\documentclass[a4paper,floatfix,rmp,twocolumn,showkeys,superscriptaddress]{revtex4}
  \usepackage[latin1]{inputenc}
  \usepackage{graphicx}
  \usepackage{mathtools}
  \setcitestyle{numbers,square,comma}

  \usepackage{amsmath}

\begin{document}

  \title{Evolutionary paths to lateralization of complex brain functions}

  \providecommand{\CNB}{Departamento de Biolog\'ia de Sistemas, Centro Nacional de Biotecnolog\'ia (CSIC), C/ Darwin 3, 28049 Madrid, Spain. }
  \providecommand{\GISC}{Grupo Interdisciplinar de Sistemas Complejos (GISC), Madrid, Spain.}
  
  \author{Lu\'is F Seoane}
    \affiliation{\CNB}
    \affiliation{\GISC}

  \vspace{0.4 cm}
  \begin{abstract}
    \vspace{0.2 cm}

    At large, most animal brains present two mirror-symmetric sides; but closer inspection reveals a range of asymmetries (in shape and function), that seem more salient in more cognitively complex species. Sustaining symmetric, redundant neural circuitry has associated metabolic costs, but it might aid in implementing computations within noisy environments or with faulty pieces. It has been suggested that the complexity of a computational task might play a role in breaking bilaterally symmetric circuits into fully lateralized ones; yet a rigorous, mathematically grounded theory of how this mechanism might work is missing. Here we provide such a mathematical framework, starting with the simplest assumptions, but extending our results to a comprehensive range of biologically and computationally relevant scenarios. We show mathematically that only fully lateralized or bilateral solutions are relevant within our framework (dismissing configurations in which circuits are only partially engaged). We provide maps that show when each of these configurations is preferred depending on costs, contributed fitness, circuit reliability, and task complexity. We discuss evolutionary paths leading from bilateral to lateralized configurations and other possible outcomes. The implications of these results for evolution, development, and rehabilitation of damaged or aged brains is discussed. Our work constitutes a limit case that should constrain and underlie similar mappings when other aspects (aside task complexity and circuit reliability) are considered. 

  \end{abstract}


\maketitle

  \section{Introduction}
    \label{sec:1}

    In Bilateria (an ample clade of animals that includes humans) the body plan displays an overall mirror-symmetric disposition. Mirror (or bilateral) symmetry captures mathematically our day-to-day experience that reflected objects look the same, but inverted side-wise (an object's left appears at the reflection's right). A mirror symmetry exists in bilaterian bodies with respect to our sagittal central plane (which separates our left and right sides). For example, both our hands look like the mirror reflection of each other with respect to that plane. 

    This symmetry is present in most parts of bilaterian central nervous systems -- including the human brain, where it also appears broken at a range of levels \cite{GalaburdaGeschwind1978, TogaThompson2003, Hugdahl2005, HerveTzourio2013, ENIGMA2018, ENIGMA2020, Seoane2020}. From a structural perspective some brain areas grow bigger than their symmetric counterpart -- for example, usually, several regions of the frontal and temporal left hemisphere are thicker than their contralateral counterpart \cite {ENIGMA2018, ENIGMA2020}. Some of these structural differences underlie an asymmetry of function as well -- human language is a preferred example as it depends on the development, in a dominant side only, of a series of areas (Broca's, Wernicke's, etc.)\ as well as exuberant connections between them around the Sylvian fissure \cite {Geschwind1972, CataniFfytche2005, FedorenkoKanwisher2009, FedorenkoKanwisher2012, FedorenkoThompson2014, BlankFedorenko2016, BerwickChomsky2016}. In some cases, functional asymmetry is observed without such salient morphological difference -- e.g.\ the right hemisphere usually dominates regarding high-level visual processing, taking care of discerning fine details; while the left hemisphere performs similar operations for rather broader scales \cite{BrownKosslyn1995, Hellige1995}. Whether linked to structure and function or not, behavior can break the mirror symmetry as well -- take hand dominance in humans, also present at different levels (and varying in choice of dominant side) in other vertebrates \cite{Galaburda1995, RogersVallortigara2004, HalpernRogers2005, Rogers2006}. 

    The discovery of Broca's area proved in one fell swoop that brain function is localized (which was unclear at the time) and that the bilateral symmetry of this organ is broken in humans. An assumption lingered among scientists that the symmetry breaking was due to the complex human cognitive abilities, thus that increased complexity would generally have a role in favoring asymmetry. Indeed, brain asymmetry was deemed a human trait that other species should lack \cite{Harrington1995, Corballis2009, MacNeilageVallortigara2009}. As this was proved wrong (and brain asymmetry was found widespread among other animals \cite{Galaburda1995, PascualPreat2004, VallortigaraBisazza1999, RogersVallortigara2004, HalpernRogers2005, Rogers2006}), neuroscientists worried less about the role of sheer complexity in symmetry breaking. They focused instead on more tangible mechanisms, such as input of light during hatching in birds \cite{HalpernRogers2005} or faster speed in single-hemisphere processing \cite{RingoSimard1994}. The hypothesis of complexity as a driving force of symmetry breaking in the brain survived with some nuances \cite{TogaThompson2003, Corballis2017}, often linked to mechanistic explanations (e.g. that that each hemisphere could have specific abilities, and more complex tasks would recruit subunits in each hemisphere differently \cite{VallortigaraBisazza1999}; or that asymmetric brains would allow a more optimal packing, thus supporting more functions \cite{Corballis2017}). But the question remains: Can cognitive complexity {\em per se} be a driving force behind the break of bilateral symmetry in neural systems? If so, is it possible to infer thresholds of complexity beyond which bilateral symmetry is doomed? Are there parsimonious ways through which the lost symmetry of a neural circuit could re-emerge? A rigorous mathematical formalism to answer these questions is lacking. 

    We tackle these issues from a computational framework, within which notions of complexity can be well defined and quantified. We assume that neural circuits are performing some computational job -- whether relaying signals, taking in inputs, or transforming them in some way. All these actions have metabolic costs (incurred into by neurons as they are engaged) but also more abstract thermodynamic costs that relate to the complexity of the computation itself or to the reliability of the operations performed \cite{KolchinskyWolpert2020, WolpertKolchinsky2020, KolchinskyWolpert2021}. 

    Reliability is important as real neural systems work within a noisy environment -- a feature itself that can be exploited for computation \cite{Maass2014, Maass2015}. How to compute with unreliable units is a problem that worried researchers since the inception of computer science \cite{vonNeumann1956, MooreShannon1956, WinogradCowan1963}. Redundancy (introducing units that perform the same operations in parallel, or that can substitute a main piece if it is damaged) is often a preferred strategy to cope with this issue. But too much redundancy would incur in unnecessary computational costs (e.g.\ as responses from parallel circuits need to be integrated) or multiply the energetic metabolic expenses. When is redundancy preferred depending on the reliability of the circuitry and the complexity of the task at hand? 

    Bilateral symmetry has been, throughout evolution, a source of redundancy that provided neural circuits in duplicated pairs. While our discussion is centered on lateralization versus bilaterality, our results are strictly general for any pair of redundant circuits -- originated in the bilateral body plan or not. As Darwinism proceeds, we expect that optimality constraints regarding computation, complexity, and redundancy will guide, allow, or prevent certain evolutionary paths. These are the possibilities that we intend to illuminate in this paper. Matters of optimality, reliable computation, and selection are also relevant during development -- as learning engages in a Darwinian process of its own \cite{Edelman1987}. Finally, the optimality of lateralized or bilateral neural configurations might be challenged again as aging proceeds \cite{OlleSole2020} or as various insults damage the nervous system. A landscape of possible, optimal configurations of neural circuits might help us navigate these cases and even suggest treatments to improve cognitive performance in damaged or aged brains. 

    \begin{figure}[] 
      \begin{center} 
        \includegraphics[width=\columnwidth]{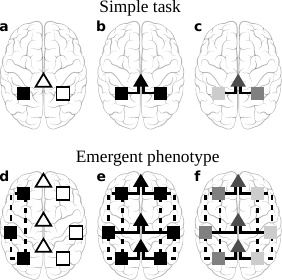}
    
        \caption{{\bf Simple models for brain bilaterality versus lateralization.} {\bf a-c} Modeling simple phenotypes. {\bf a} A simple task can be carried out by a single unit (in this case, in the left hemisphere -- black square). This would be a lateralized configuration. A similar unit at the right hemisphere (white square) and some circuit or mechanism to integrate the activity of both symmetric counterparts (white triangle) are left unused. {\bf b} Instead, both mirror-symmetric units might be engaged to solve the simple task (black squares). In this case, the mechanism that integrates both outputs (black triangle) might be needed. This is a bilaterally-symmetric configuration. {\bf c} It might be possible to implement the task by engaging both units in some gradual manner (partial recruitment is marked by shades of gray). That would demand a graded use of the integrating mechanism as well (shaded triangle). {\bf d-f} Modeling an emerging phenotype that recruits $K$ (in this case, $3$) different modules, each implementing a simple task. Regarding lateralized ({\bf d}) versus full ({\bf e}) or partial ({\bf f}) bilateral solutions, we find the same possibilities as before. What configurations are optimal in each case? }
    
        \label{fig:1}
      \end{center}
    \end{figure}

    In this paper we lay out a minimal mathematical model to map optimal configurations (bilateral versus fully lateralized -- allowing any intermediate designs) of computing neural units as a function of their running costs, reported benefits, reliability (as measured by an error rate), and complexity of the task at hand. All mathematical details are developed in the appendices. In the Results section we discuss our most important insights. In Sec.\ \ref{sec:3.1} we study the least complicated case, in which a simple task (so simple that it cannot be decomposed further) is implemented either by a faulty, irreducible neural unit, or by that faulty unit and its mirror symmetric counterpart (Fig.\ \ref{fig:1}{\bf a-c}). While any combination of engagement of either unit is allowed, we show that only fully lateralized (Fig.\ \ref{fig:1}{\bf a}) or fully bilateral (Fig.\ \ref{fig:1}{\bf b}) solutions matter. In Sec.\ \ref{sec:3.2} we study a complex computation that needs the cooperation of several such couples of units, with each bilaterally symmetric couple taking care of a strictly different subtask (Fig.\ \ref{fig:1}{\bf d-f}). We say that this is an emergent or complex computation, or an emergent or complex phenotype of the cooperating couples of units. We asses under what conditions such phenotypes might emerge, and whether they entail partial (Fig.\ \ref{fig:1}{\bf f}) or full lateralization or bilateral configurations (finding, again, that only all-or-nothing engagement is relevant; Fig.\ \ref{fig:1}{\bf d-e}). In Sec.\ \ref{sec:3.3} we use our model to explore evolutionary or developmental paths that should lead from a bilateral to a fully lateralized configuration. While we use the simplest possible model to guide our discussion, we prove in the appendixes that some results are quite general. An ample range of biologically and computationally meaningful choices of costs and fitness functions should share the most salient features that we derive -- notably, that only fully lateralized or completely bilateral configurations matter. In the Discussion, we review some real neural structures using our framework and highlight the implications of our results for development and treatment in damaged, diseased, or aged brains.


  \section{Results}
    \label{sec:3}

    \subsection{Charting bilaterality and lateralization for simple tasks}
      \label{sec:3.1}

      \begin{figure*}[] 
        \begin{center} 
          \includegraphics[width=\textwidth]{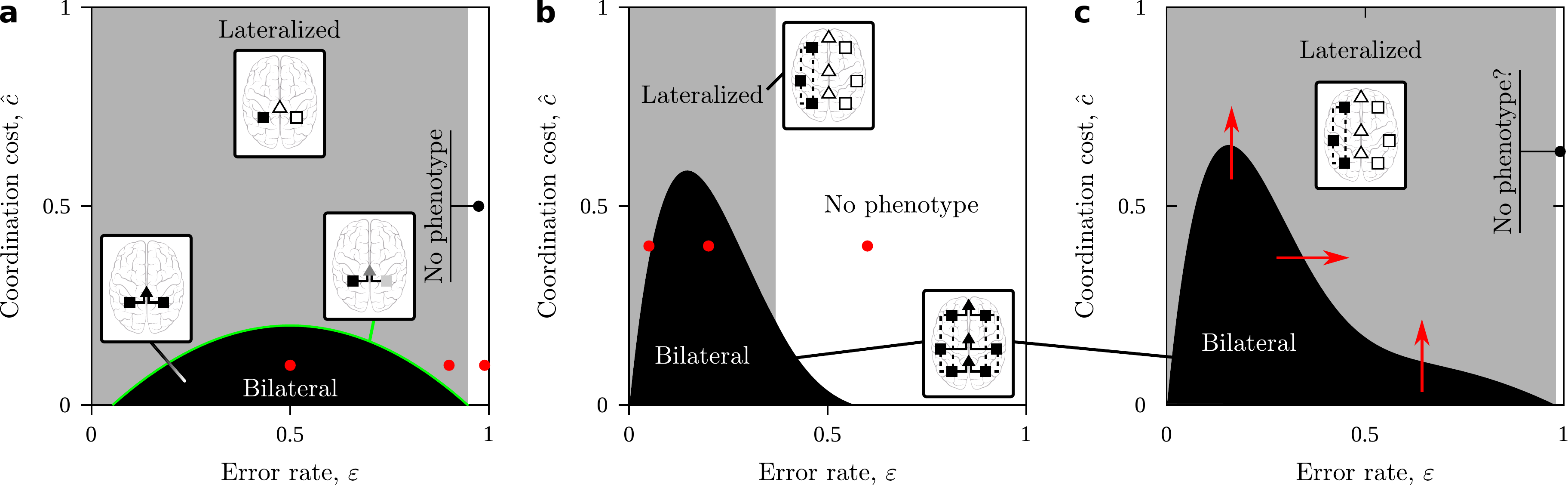}
      
          \caption{{\bf Optimal lateral vs bilateral configurations.} For $(\varepsilon, \hat{c})$ within white areas, the sought phenotype is so costly that it never pays off. Within light gray regions, it is optimal to lateralize. Within black areas, bilaterality is preferred. {\bf a} Model for simple cognitive tasks with $b=1$, $c = 0.05$. Such simple computations are the only ones in which we find graded solutions (green curve separating lateralized and bilateral configurations), but they occupy a negligible part of the phase space. Red dots correspond to the utility functions shown in Fig.\ \ref{fig:app1.1}. {\bf b} Model for strictly emergent phenotypes with $\tilde{b}=1$, $c=0.01$, $K=10$. Red dots correspond to the utility functions shown in Fig.\ \ref{fig:app2.1}. {\bf c} Model with emergent phenotypes upon a substrate that retains the simpler functionality with $b=1$, $\tilde{b}=1$, $c=0.05$, $K=10$. Arrows indicate trajectories across parameter space prompted by degradation of the neural substrate, which might increase coordination costs (vertical arrows) or fallibility (horizontal arrow). }
      
          \label{fig:2}
        \end{center}
      \end{figure*}

      Assume that some neural circuit has to solve a relatively simple task. Assume also that, due to our evolutionary history and bilaterian body plan, we posses two such circuits: one at the left, $L$, and one at the right, $R$ (squares in Fig.\ \ref{fig:1}{\bf a-c}). We will refer to them as left and right units or circuits, and we will say that they conform a mirror-symmetric or bilateral couple of units or circuits. Further assume that they are faulty such that with a probability $\varepsilon$ they fail each time that they attempt their computation. 

      Does it pay off to keep both mirror-symmetric circuits working? Note that two units together might perform the task more reliably. However, keeping each circuit running has a metabolic cost -- as neurons need to be fed with blood when they are active. Also, if both units function simultaneously, they might interfere with each-other -- one side providing a wrong answer might spoil the other's correct outcome. Some additional structure or mechanism is needed to cross-check simultaneous activity (triangles in Fig.\ \ref{fig:1}{\bf a-c}). Let us attempt to capture, with the simplest equation possible, all the costs and benefits of keeping just one unit of the mirror symmetric couple running (Fig.\ \ref{fig:1}{\bf a}) versus sustaining both circuits engaged (Fig.\ \ref{fig:1} {\bf b}) versus keeping some intermediate level of activity (Fig.\ \ref{fig:1}{\bf c}), or even shutting both of them all together. 

      Say that, whenever computing this task is required, unit $L$ is switched on with a probability $l \in [0,1]$ and unit $R$ is activated with a probability $r \in [0,1]$. If we deal with a computation that needs to be carried out uninterrupted throughout the day, we can think of $l$ and $r$ as the fraction of time that each unit remains active. We introduce a cost $c$ paid for each occasion in which either unit is switched on, such that 
        \begin{eqnarray}
          C &=& c(l+r) 
          \label{eq:3.1.1}
        \end{eqnarray}
      are the total expenses of running both units independently. This stands for some metabolic cost of use. Note that the mere structural existence of each unit should have a great cost as well -- but that is paid independently of use, and only once as the brain develops. Adding such additional cost does not alter our results qualitatively. We could also think of costs that depend, e.g., on the desired accuracy, such that units that compute with a lower error are more costly. One such case was explored partially in \cite{Seoane2020}. In App.\ \ref{app:5} we show that the generality of our results largely includes this scenario as well. 

      The coordinating structure or mechanism has a cost $\hat{c}$ of its own, and we assume that it is paid whenever both bilateral units are functioning simultaneously. Hence: 
        \begin{eqnarray}
          \hat{C} &=& \hat{c}lr
          \label{eq:3.1.2}
        \end{eqnarray}
      are the total coordination expenses. This can capture the metabolic expenses of the coordinating structure or mechanism, but also any losses due to interference between both units that results in a faultier functioning. We can think of this last possibility as contributing an average loss due to insufficient coordination. We could split these costs, but each of them is contributed only whenever both units are active -- thus they can all be absorbed within $\hat{c}$. 

      As for the benefits, we assume an all-or-nothing scenario such that some fitness $b$ is gained if and only if the task is successfully implemented. We could think of alternatives -- e.g.\ even an imperfect implementation of the task contributes some fitness. Such possibilities are explored in App.\ \ref{app:5}. Focusing on the simplest Ansatz, we get: 
        \begin{eqnarray}
          B &=& b\left[ (1-\varepsilon)l(1-r) + (1-\varepsilon)(1-l)r + (1-\varepsilon^2)lr \right]. \nonumber \\
          \label{eq:3.1.3}
        \end{eqnarray}
      The first term is the probability that $L$ is active, $l$, and works properly, $1-\varepsilon$, and that $R$ is switched off, $1-r$. The second term is the probability that $R$ is active and works properly and that $L$ is switched off. The third term is the probability that both units are active and at least one of them produces the correct answer (the probability that neither does is $\varepsilon^2$). In this simplest Ansatz we assume that the correct answer cannot be reached if neither unit did on its own. We could, alternatively, assume that two faulty but nearly correct answers could improve each-other (especially thanks to the coordinating mechanism). To capture this possibility, we would use, in this third term, an alternative function of $\varepsilon$ instead of $1-\varepsilon^2$. This is explored in App.\ \ref{app:5}. Back to our simplest equation, what this third term assumes is that, thanks to the coordinating mechanism, if the correct outcome has been produced by at least one unit, it can always be picked up. This does not consider interference as discussed above; but then again, that could be captured within $\hat{C}$. 

      Subtracting costs from benefits renders a utility function: 
        \begin{eqnarray}
          \rho(l,r) &\equiv& B - C - \hat{C} \nonumber \\ 
          &=& b(1-\varepsilon)\left[ l(1-r) + (1-l)r + (1+\varepsilon)lr \right] \nonumber \\ 
          && -c(l+r) - \hat{c}lr. 
          \label{eq:3.1.4}
        \end{eqnarray}
      Given some values to our model parameters ($b$, $c$, $\hat{c}>0$ and $\varepsilon \in [0,1)$) the maximum utility as a function of $l$ and $r$ tells us the optimal configuration of our units -- i.e.\ which level of activity should be kept in either mirror-symmetric circuit. 

      We can draw our optimal configurations in a map (or morphospace) to chart how they change as we vary our parameters. Fig.\ \ref {fig:2}{\bf a} shows such a map for Eq.\ \ref{eq:3.1.4} for fixed $b=1$ and $c=0.05$ and varying $\varepsilon \in [0,1)$ and $\hat{c} \in [0,1]$. We observe a white, vertical stripe for $\varepsilon > 1 - c/b$ in which both units are so faulty that the fitness contributed by this task does not pay off enough to keep any circuit active. As such, the phenotype resulting from implementing the task at hand does not appear. In the large gray area, coordination is costly enough that it pays off to lateralize and keep only one unit (indistinctly $L$ or $R$) and shut off the other one. In the smaller black area, it is always convenient to keep both units engaged. At the boundary separating both regions (green curve), the optimal solution has one unit always active and the other one active any arbitrary fraction of the time. This is the only graded configuration (in which both units are not either completely off or completely on) -- otherwise, Eq.\ \ref{eq:3.1.4} has only all-or-nothing solutions. 

      This is the general shape of the morphospace for Eq.\ \ref{eq:3.1.4} for any parameters (see App.\ \ref{app:1} for its mathematical derivation). The boundary between bilateral and lateralized solutions is a parabola with a maximum of $\hat{c} = b/4 - c$ located at $\varepsilon=1/2$. Note that if $c > b/4$, the bilateral configuration disappears (Fig.\ \ref{fig:app1.2}{\bf c}), so that the lateralized solution is preferred for any $ (\varepsilon, \hat{c})$. If $c=0$, the maximum of the parabola is at $b/4$. This imposes a very stringent limit: coordination costs can never be larger than a fourth of the contributed fitness.

    \subsection{Charting bilaterality and lateralization for emergent phenotypes}
      \label{sec:3.2}

      Conceive now a complex task that, in order to be successfully implemented, needs to recruit a series of brain regions, each one carrying out a specific and different subtask. A good example is the human language ability, that requires the successful functioning of Broca's and Wernicke's areas as well as retrieving information from a semantic map, etc.\ \cite{Geschwind1972, CataniFfytche2005, FedorenkoKanwisher2009, FedorenkoKanwisher2012, FedorenkoThompson2014, BlankFedorenko2016, BerwickChomsky2016}. Failure at any of these subtasks results in specific pathology related to the malfunctioning area. Full-fledged language only emerges if all regions perform correctly. Let us call such computations (that require different units to collaborate) {\em emergent} phenotypes. Assuming that each subtask can be implemented as before (i.e. by either unit within a mirror-symmetric couple), then we ask again: When is it favorable to lateralize and keep just one side active (Fig.\ \ref {fig:1}{\bf d}), keep all bilateral couples functioning (Fig.\ \ref{fig:1}{\bf e}), have them running at some intermediate level (Fig.\ \ref{fig:1}{\bf f}), or switch the whole circuit off completely and, consequently, fail to implement the complex computation? 

      Let us assume an emergent phenotype that consists of $K$ subtasks. We will take $K$ as a proxy for the cognitive complexity of this phenotype. Let us assume that each of these subtasks can be implemented by a couple of mirror-symmetric units as the ones discussed above, and that each of these couples incurs separately in the same running and coordination costs, such that: 
        \begin{eqnarray}
          C &=& cK(l+r), \nonumber \\ 
          \hat{C} &=& \hat{c}Klr. 
          \label{eq:3.2.1}
        \end{eqnarray}
      As before,  we can interpret $l$ and $r$ as the average time that units at the left or right sides are active. Alternatively, we can say that a fraction $l$ of the $K$ left units is always switched on (and similarly for the right side). We have assumed, without loss of generality, that all subtasks are equally costly -- if not, we could make $c$ and $\hat{c}$ the corresponding averages. 

      Regarding the fitness benefit of the emergent phenotype, we insist that it is only fully cashed in if all independent subtasks are successfully implemented, thus: 
        \begin{eqnarray}
          \tilde{B} &=& \tilde{b}K \left[ (1-\varepsilon)l(1-r) + (1-\varepsilon)(1-l)r \right. \nonumber \\ 
          && \left. + (1-\varepsilon^2)lr \right]^K. 
          \label{eq:3.2.2}
        \end{eqnarray}
      Here we see the same probability of implementing each subtask as before -- now raised to the $K$-th power, which gives us the likelihood that no subtask is lacking. We assume that the total benefit reported by the emergent phenotype is $\tilde{b}K$ (we could absorb the $K$ within $\tilde{b}$, but it is convenient not to). As before, in App.\ \ref{app:5} we show that the following results are more general if we would choose different dependencies on $\varepsilon$. 

      We can now define the following utility function: 
        \begin{eqnarray}
          \rho(l,r) &\equiv& \tilde{B}/K - C/K - \hat{C}/K \nonumber \\ 
          &=& \tilde{b}(1-\varepsilon)^K\left[ l(1-r) + (1-l)r + (1+\varepsilon)lr \right]^K \nonumber \\ 
          && - c(l+r) - clr. 
          \label{eq:3.2.3}
        \end{eqnarray}
      Fig.\ \ref{fig:2}{\bf b} charts its optimal solutions with $\tilde{b}=1$, $c=0.01$, $K=10$, and varying $\varepsilon \in[0,1)$ and $\hat{c} \in [0,1]$ (see App.\ \ref{app:2} for the mathematical derivation of this figure). 

      The area in which the phenotype fails to emerge (now found for $\varepsilon > 1 - \sqrt[K]{c/\tilde{b}}$, white region in Fig.\ \ref{fig:2}{\bf b}) is much wider than in the previous case. Meanwhile, the combination of parameters for which the lateralized configuration is optimal (gray) has shrunk. Note that we have taken a relatively low running cost for independent circuits ($c = 0.01$). If these costs were even lower ($c \rightarrow 0$), the region in which the phenotype fails to emerge would become negligible as $\varepsilon > 1 - \sqrt[K]{c/\tilde{b}} \rightarrow 1$. This situation is noteworthy, even though we expect realistic scenarios to have non-negligible costs ($c > 0$). Even if we do not approach the $c \rightarrow 0$ limit, we can expect that complex phenotypes contribute a much larger fitness than the implementation of simpler tasks -- think, e.g., the advantages brought about by human language. In that case, since we set $\tilde{b}=1$ to generate our maps, we should re-scale the running costs accordingly, resulting in much smaller $c$. We revisit this possibility in Sec.\ \ref{sec:3.3.1}. 

      The region in which the bilateral combination is optimal is shifted to lower values of $\varepsilon$ and deformed with respect to the parabola. Unlike before, along the boundary it is not optimal to keep one side switched on and the other side active to an arbitrary degree. Instead, at the boundary between the black and gray regions, both the lateralized and fully bilateral configurations (but no others) are indifferently optimal. Thus now we find even less graded solutions than for simple tasks: all optimal configurations are either both units switched off, on, or just one completely active side. 

      It can be proved (see App.\ \ref{app:2}) that the curve separating lateralized and bilateral configurations has only one maximum, so all maps have a similar shape as we vary our parameters. As the complexity of the emergent task increases (i.e.\ as $K$ grows because more distinct functionality needs to be recruited), the area in which the phenotype fails to emerge grows (as per the condition $\varepsilon > 1 - \sqrt[K]{c/\tilde{b}}$). The region of bilaterality shifts further left while its peak reaches higher in the $\hat{c}$ axis. For simple tasks there was a harsh limit ($\hat{c} < b/4$) for the optimality of the bilateral configuration. For complex, emergent phenotypes, a much higher coordination cost can be tolerated while the mirror symmetric solution is sustained. \\ 

      So far we have described the very unlikely scenario in which a complex phenotype emerges very swiftly, fully recruiting all the needed units for its own good. Evolutionarily, this resembles a hopeful monster, unlikely in a gradual Darwinian framework. We shall refer to such case as a {\em strictly emergent} phenotype. More realistically, complex cognition will engage the needed tasks in a progressive fashion. While not recruited, the corresponding units can keep implementing their former jobs. In such a case we should add up the sustained fitness contribution brought about by the earlier, independently implemented phenotypes. 

      Since we have $K$ tasks (which, for simplicity, we assume each one reports a fitness benefit as in Eq.\ \ref {eq:3.1.3}), we can write the following utility function: 
        \begin{eqnarray}
          \rho(l,r) &\equiv& \tilde{B}/K + KB/K  - C/K - \hat{C}/K \nonumber \\ 
          &=& \tilde{b}(1-\varepsilon)^K\left[ l(1-r) + (1-l)r + (1+\varepsilon)lr \right]^K \nonumber \\ 
          && + b(1-\varepsilon)\left[ l(1-r) + (1-l)r + (1+\varepsilon)lr \right] \nonumber \\ 
          && - c(l+r) - clr. 
          \label{eq:3.2.4}
        \end{eqnarray}
      Fig.\ \ref{fig:2}{\bf c} charts its optimal configuration for $\tilde{b}=1$, $b=0.5$, $c=0.05$, and $K=10$ (see App.\ \ref{app:3} for the mathematical details). 

      Compared to the case with the strictly emergent phenotype, the region in which the complex phenotype combined with the ancient, independent ones fails to emerge is much reduced (white stripe at the right with $\varepsilon > 1 - b/c$; but the condition is actually more lenient -- see App.\ \ref{app:3}). The areas for both lateralized (gray) and bilateral (black) configurations have expanded. Thus the existence of an evolutionary substrate that remains operative largely enables the process through which a complex phenotype can come into existence. 

      As before, only all-or-nothing solutions are observed -- also along the boundary between lateralized and bilateral configurations. The curve tracing this boundary looks like a mixture of the parabola (from the simplest scenario) and the peaked separation between configurations for strictly emergent phenotypes. The separation curve now might present one or two peaks. The higher peak is always present, and located at smaller error rates. As before, it grows above $\hat{c} = \tilde{b}/4$, allowing bilateral solutions with much larger coordination costs.

    \subsection{Evolutionary paths}
      \label{sec:3.3}

      The morphospaces derived above give us static pictures: what is and what is not optimal or feasible given some fixed conditions? But evolution is a dynamic process and we are interested in what pathways might allow (or force) us to transit from one configuration to another, or when does a given design remain optimal as surrounding constraints change. By combining the maps above we can see how a configuration becomes optimal or suboptimal as a complex phenotype emerges out of simpler ones. 

      \subsubsection{Swift emergence of complex phenotypes}
        \label{sec:3.3.1}

        \begin{figure*} 
          \begin{center} 
            \includegraphics[width=\textwidth]{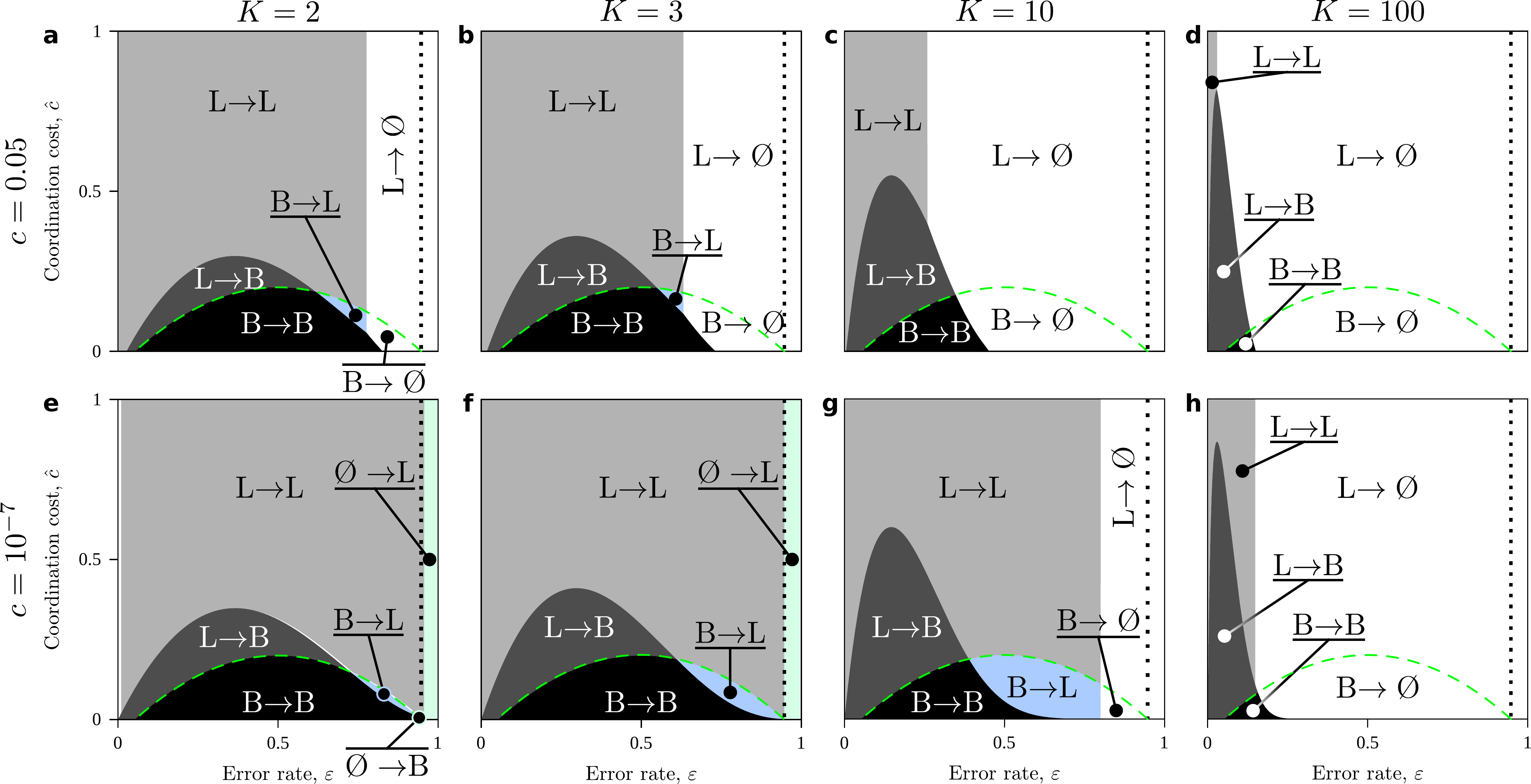}
        
            \caption{{\bf Evolutionary paths to strictly emergent phenotypes.} Taking $b=1$: {\bf a-d} Rather large cost of running units independently ($c=0.05$). Different evolutionary paths (labeled as explained in the main text) become available or are lost depending on the complexity of the emergent phenotype: {\bf a} $K=2$, {\bf b} $K=3$, {\bf c} $K=10$, and {\bf d} $K=100$. {\bf e-h} With negligible running costs ($c=10^{-7}$), new pathways (green, labeled {\bf \O$\boldsymbol{\rightarrow}$L} and {\bf \O$\boldsymbol{\rightarrow}$B}) are present when the complexity of the emergent phenotype is not too large ({\bf e} $K=2$ and {\bf f} $K=2$), but not so for more complex emerging tasks ({\bf g} $K=10$ and {\bf h} $K=100$). }
        
            \label{fig:3}
          \end{center}
        \end{figure*}

        Let us look first at the swift emergence of complex phenotypes -- a scenario compared to hopeful monsters above. Fig.\ \ref{fig:3} shows the overlap between the morphospace for simple tasks and the morphospace for strictly emergent phenotypes. The former is represented by the dashed green curve (boundary between bilateral and lateralized solutions from Fig.\ \ref{fig:2}{\bf a}) and by a dotted vertical black line at $\varepsilon = 1-c/b$ (beyond which the simple task is too costly to assume). 

        In all conditions reported in Fig.\ \ref{fig:3}, we observe (often broad) regions marked {\bf B$\boldsymbol{\rightarrow}$B} (black) and {\bf L$\boldsymbol{\rightarrow}$L} (light gray) in which, respectively, the bilateral and lateralized solutions are optima for the implementation of both the simple and complex phenotypes. We also find ample regions marked {\bf B$\boldsymbol{\rightarrow}$\O} and {\bf L$\boldsymbol{\rightarrow}$\O} (white, Fig.\ \ref{fig:3}{\bf a-d} and {\bf g-h}) in which there was an optimal configuration for the simple task, but in which a profitable circuit fails to emerge for the complex phenotype. This is due to the condition $\varepsilon < 1- \sqrt[K]{c/\tilde{b}}$, which is usually more stringent than $\varepsilon < 1-c/b$. But, as discussed above, this situation can be reversed: if the emergent phenotype is much more profitable than the simple one, and we re-scale costs accordingly, we can find $1- \sqrt[K]{c/\tilde {b}} \rightarrow 1$. Then, we might observe {\bf \O$\boldsymbol{\rightarrow}$L} (green, Fig.\ \ref{fig:3} {\bf e-f}) and {\bf \O$\boldsymbol{\rightarrow}$B} (tiny region at the bottom right part in Fig.\ \ref{fig:3}{\bf e}), in which it is never favorable to implement the simple task but the complex one is so valuable that an optimal, functioning circuit emerges. While mathematically favored, the evolution of such complex circuits de novo might be biologically unfeasible. 

        For all conditions shown in Fig.\ \ref{fig:3} we see salient regions labeled {\bf L$\boldsymbol{\rightarrow}$B} (dark gray). In this interesting scenario, a lateralized solution is optimal for carrying out the simple task, but a bilateral configuration would be preferred for the emergent phenotype. Depending on an organism's evolutionary history, the symmetric counterpart of a lateralized circuit might have been lost (or devoted to other tasks -- see below), hence recovering the bilateral configuration might not be possible. In that case, the organism might get stuck with the suboptimal, lateralized solution -- a {\em frozen accident}. Alternatively, the evolutionary pressure towards a redundant solution could foster the appearance of a duplicate. This duplicate does not need to be a bilateral counterpart -- as noted above, we label our units {\em left} and {\em right} to focus the discussion on mirror symmetry, but our results are valid for any set of duplicated circuits. The evolution of such duplicates has been observed in the mammalian brain \cite{ChakrabortyJarvis2015}. Our {\bf L$\boldsymbol{\rightarrow}$B} regions suggest ample pressures favoring this evolutionary pathway. 

        In Figs.\ \ref{fig:3}{\bf a-b} and {\bf e-g} we find regions marked {\bf B$\boldsymbol{\rightarrow}$L} (blue). In them, the bilateral symmetry that is optimal to solve the simple task is broken as a more complex phenotype (as measured by the number, $K$, of different subunits recruited) takes over. The sheer complexity of the emergent phenotype forces this symmetry breaking, proving that computational complexity can be a driving force behind the lateralization of advanced cognition. 

        This {\bf B$\boldsymbol{\rightarrow}$L} region vanishes for larger $K$ -- i.e. for more complex phenotypes, which recruit more mirror-symmetric modules. That happens because larger $K$ move the threshold $\varepsilon = 1 - \sqrt[K]{c/\tilde{b}}$ (beyond which the phenotype does not pay off) towards lower values of $\varepsilon$. This prevents the assembly of faulty circuits to implement the complex task. Note that the overlap between bilateral solutions for the simple and complex phenotypes ({\bf B$\boldsymbol{\rightarrow}$B}) also shrinks as $K$ grows. Thus, if running costs are kept high ($c \gg 0$), the original mirror-symmetric solution tends to disappear when a complex phenotype emerges swiftly (e.g.\ Fig.\ \ref{fig:3}{\bf d}). As before, this is alleviated if $c \rightarrow 0$ (when re-scaled by $\tilde{b}$, Figs.\ \ref{fig:3}{\bf e-h}). This scenario sustains broader {\bf B$\boldsymbol{\rightarrow}$L} and {\bf B$\boldsymbol{\rightarrow}$B} regions, but they also dwindle for very complex emergent phenotypes (Fig.\ \ref{fig:3}{\bf h}).

      \subsubsection{Complex phenotypes emerging upon circuits for simpler tasks} 
        \label{sec:3.3.2} 

        \begin{figure*}[] 
          \begin{center} 
            \includegraphics[width=\textwidth]{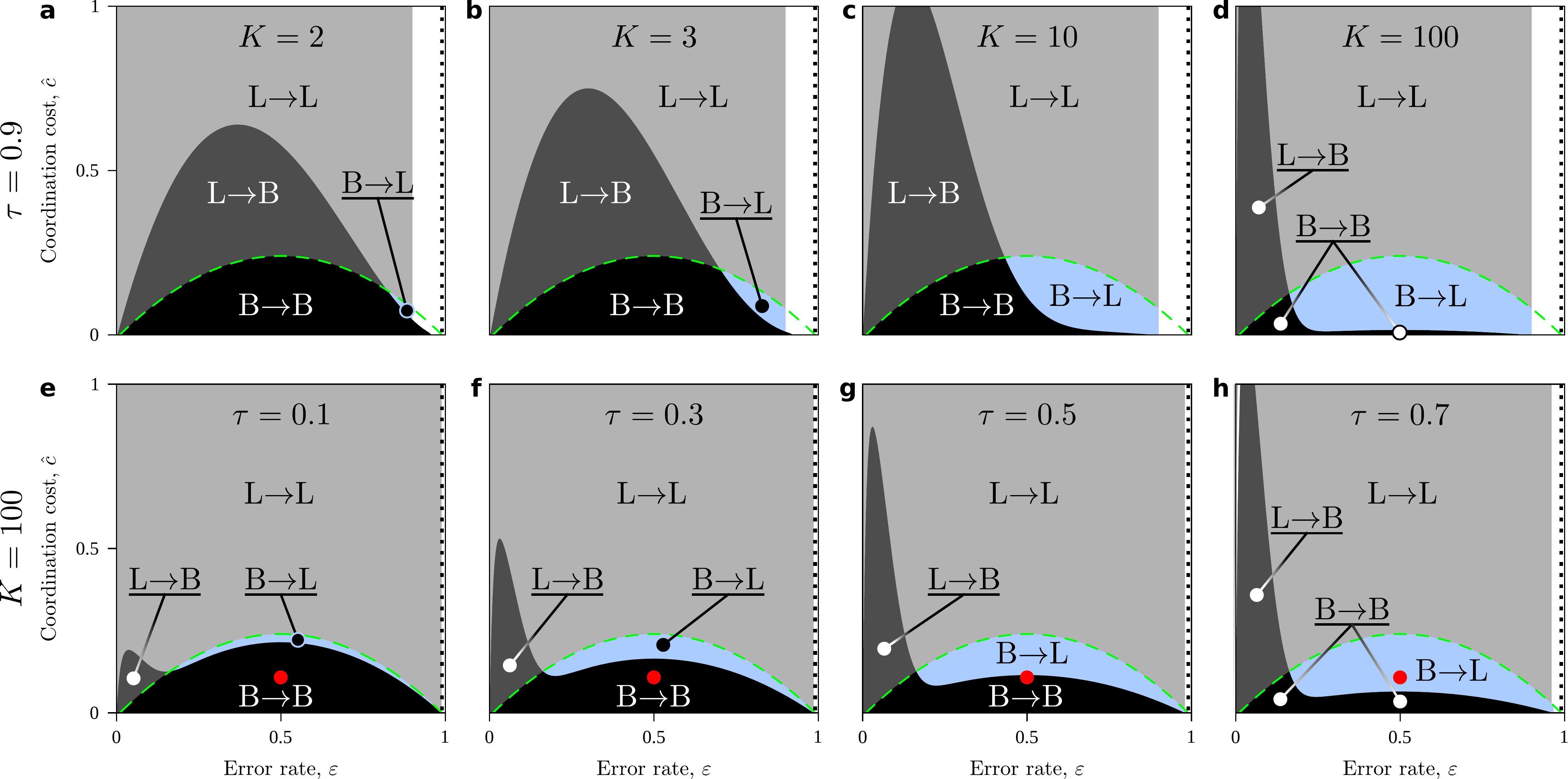}
        
            \caption{{\bf Evolutionary paths to complex phenotypes that emerge upon substrates that retain their simpler functions.} With $b=1$, $\tilde{b}=2$, and $c=0.01$ in all cases. The pathway from bilateral to lateralized configurations {\bf B$\boldsymbol{\rightarrow}$L} is more robust now. {\bf a-d} We explore an emergent phenotype that occupies the neural substrate $90\%$ of the time ($\tau = 0.9$). As the complexity of the emergent phenotype increases ({\bf a} $K=2$, {\bf b} $K=3$, {\bf c} $K=10$, {\bf d} $K=100$), it becomes unavoidable that the mirror symmetry breaks apart. {\bf e-h} We explore a notably complex phenotype ($K = 100$) that gradually increases the fraction of time during which it makes use of the neural substrate ({\bf e} $\tau = 0.1$, {\bf f} $\tau = 0.3$, {\bf g} $\tau = 0.5$, {\bf h} $\tau = 0.7$; with {\bf d} $\tau = 0.9$ completing the progression). This resemble developmental situations in which higher brain functions are assembled gradually and displace simpler computations in a same neural substrate. A circuit sitting where the red dot is would become lateralized by such process. }
        
            \label{fig:4}
          \end{center}
        \end{figure*}

        Complex phenotypes are likely to evolve upon a previously existing substrate. This would consist of couples of mirror symmetric units -- each couple already implementing its own, individual, simpler task. Most likely, those original functions are not completely displaced while the emergent phenotype comes into being. Let us assume that, throughout the day, the emergent task engages its circuitry a fraction $\tau \in [0,1]$ of the time, and that the simpler tasks can make use of their units during the remaining time, $1 - \tau$. Assuming that the emergent and simple phenotypes contribute a fitness $\tilde{b}'$ and $b'$ respectively, we can substitute $\tilde{b} = \tau\tilde{b}'$ and $b = (1-\tau)b'$ in equation \ref{eq:3.2.4}. Note that if $\tau = 1$ we recover the case just discussed in which the emergent phenotype displaces the ancient ones completely. 

        When we superpose the resulting morphospaces, the picture changes notably from that of strictly emergent phenotypes. Fig.\ \ref{fig:4}{\bf a-d} shows evolutionary paths for an emergent task with a fitness contribution that is not outstanding ($\tilde{b}'=2$ versus $b'=1$) when $\tau=0.9$ (i.e.\ simpler tasks are only present $10\%$ of the time). This tangential yet sustained presence of simpler tasks shrinks notably the combinations of parameters for which emergent phenotypes are not viable -- compare the white areas in Fig.\ \ref{fig:3} with those in Fig.\ \ref{fig:4}. Regions in which the lateralized solution is optimal are much expanded. They now invade profusely the area of bilateral solutions for simpler phenotypes (marked in blue and labeled {\bf B$\boldsymbol{\rightarrow}$L}, in the figure). 

        This constitutes a more solid route to the lateralization of mirror-symmetric circuits due to the sheer complexity of the emergent task. Note that, in order to observe large {\bf B$\boldsymbol{\rightarrow}$L} regions in Fig.\ \ref{fig:3}{\bf e-h}, we needed to re-scale the costs by five orders of magnitude ($c=0.05$ in Fig.\ \ref{fig:3}{\bf a-d} versus $c=10^{-7}$ in Fig.\ \ref{fig:3}{\bf e-h}). This comes about because we assume that the fitness benefit from the emergent phenotype is $5$ orders of magnitude larger. Instead, in Fig.\ \ref{fig:4} the complex task contributes only twice as much. This means that the mere retention of the old phenotype (which, evolutionarily, is the more parsimonious pathway) strongly enables the emergence of complex phenotypes. These, in turn, would more likely cause the lateralization of brain activity. 

        Fig.\ \ref{fig:4}{\bf e-h} shows how the available evolutionary paths change as we increase the fraction of time devoted to the complex phenotype. This can be important to discuss function lateralization during development: as a person matures, higher brain functions are more likely to be engaged for longer time periods, potentially displacing simpler tasks. We observe broad regions in which bilateral circuits should lateralize as the more complex phenotype takes over. 

        We still observe broad regions {\bf L$\boldsymbol{\rightarrow}$B}, which constitute pressures for the evolution of redundant circuitry prompted by the emergence of a complex phenotype.

      \subsubsection{Segregating functions}
        \label{sec:3.3.3} 

        \begin{figure*}[] 
          \begin{center} 
            \includegraphics[width=\textwidth]{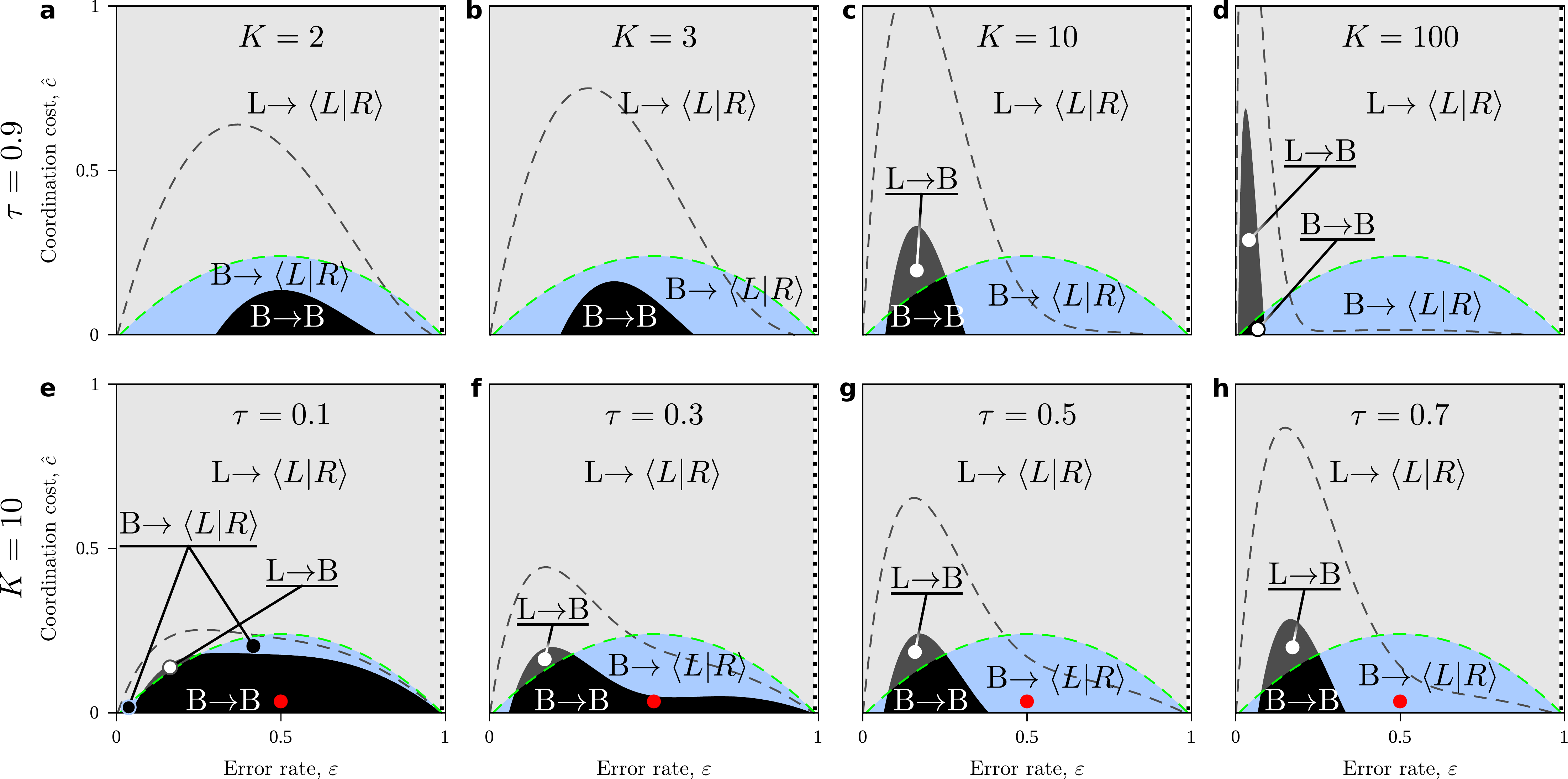}
        
            \caption{{\bf Evolutionary paths to emerging phenotypes when simpler and complex tasks can be segregated to different hemispheres.} With $b=1$, $\tilde{b}=2$, and $c=0.01$ in all cases. Gray dashed lines represent the boundary between bilateral and lateralized solutions from Fig.\ \ref{fig:4}. {\bf a-d} As before, we explore emerging phenotypes that occupy the neural substrate $90\%$ of the time ($\tau = 0.9$). Again, as the complexity of the emerging phenotype increases ({\bf a} $K=2$, {\bf b} $K=3$, {\bf c} $K=10$, {\bf d} $K=100$), it becomes unavoidable that bilateral circuits lose their mirror symmetry. This time, when bilaterality is lost, each mirror symmetric counterpart becomes specialized in either the simple or complex tasks ({\bf B$\boldsymbol{\rightarrow\left<L|R\right>}$}). We explore a moderately complex phenotype ($K = 10$) that gradually increases the fraction of time during which it makes use of the neural substrate ({\bf e} $\tau = 0.1$, {\bf f} $\tau = 0.3$, {\bf g} $\tau = 0.5$, {\bf h} $\tau = 0.7$; with {\bf c} $\tau = 0.9$ completing the progression). Again, this serves us to model a developmental situations in which higher brain functions are assembled gradually and displace simpler computations in a same neural substrate. A circuit sitting where the red dot is would become lateralized by such process. }
        
            \label{fig:5}
          \end{center}
        \end{figure*}

        In the configurations explored in the previous section, both the emergent and simple phenotypes are implemented by the same circuits -- whether mirror-symmetric or lateralized. Alternatively, both sets of tasks can be lateralized and segregated, confining each to a different hemisphere that would become specialized. When does this configuration pay off? 

        Fig.\ \ref{fig:5} shows evolutionary pathways towards the lateralized-segregated solution (noted $\left<L|R\right>$, see App.\ \ref{app:4} for mathematical details). This is amply preferred both as the complexity, $K$, of the emergent phenotype increases (Fig.\ \ref{fig:5}{\bf a-d}) and as the fraction, $\tau$, of time devoted to the complex task grows (Fig.\ \ref{fig:5}{\bf e-h}). In Fig.\ \ref{fig:5}, the boundaries between bilateral and lateralized solutions from Fig.\ \ref{fig:4}, as well as the features from the morphospace of simple tasks, have been left as references (respectively shown as dark gray dashed and green dashed curves). 

        The pathway {\bf L$\boldsymbol{\rightarrow\left<L|R\right>}$} (light gray) departs from a structure that is already lateralized for the implementation of simple tasks. This earlier lateralization might have left the mirror-symmetric structures unused, thus available for the emergent phenotype to recruit. On the other hand, those circuits might have been lost over the course of evolution -- hence this pathway would not be straightforwardly available. Again, since our results are valid for sets of duplicated circuits (not only mirror-symmetric units), the optimality of this segregated solution might be an evolutionary pressure towards the duplication of existing structures within one hemisphere (as discussed in \cite{Seoane2020, ChakrabortyJarvis2015}). 

        The {\bf B$\boldsymbol{\rightarrow\left<L|R\right>}$} pathway (blue) is more amply preferred the more complex the emergent phenotype is (i.e.\ the larger $K$ grows). This shows, again, how computational complexity can be a very strong evolutionary driver of lateralization -- and (since our results are more general) of symmetry breaking in the brain. 

        In the literature it is discussed how segregating function can be very convenient -- e.g., to have specialized hemispheres that complex cognition can recruit units from \cite{VallortigaraBisazza1999} or to allow more efficient packing \cite{Corballis2017}. Notwithstanding, even if allowing segregation, the bilateral configuration is optimal in ample regions of parameter space. Note that segregation would not happen in the bilateral solution, since it requires both mirror-symmetric circuits engaged for either the simple or complex phenotypes. The persistence of bilateral designs enable the pathways {\bf B$\boldsymbol{\rightarrow}$B} (black in Fig.\ \ref{fig:5}) and {\bf L$\boldsymbol{\rightarrow}$B} (dark gray, Fig.\ \ref{fig:5}{\bf c-h}). In these cases, the gained fitness from combining faulty circuits into robust ones can overcome the advantages of segregating tasks into equally faulty circuits. The {\bf L$\boldsymbol{\rightarrow}$B} pathway again constitutes a pressure towards duplication of existing circuits or (if still available) re-recruiting the formerly lateralized mirror-symmetric counterpart.

  \section{Discussion}
    \label{sec:4}

    Since very early in the history of computer science, redundancy was acknowledged as an efficient strategy to enable computation with faulty parts \cite{vonNeumann1956, MooreShannon1956, WinogradCowan1963}. The bilaterian body plan is a source of redundancy for many organs, including the central nervous system. Specifically, the brain is equipped with mirror-symmetric duplicates of most cortical regions, ganglia, and other subsystems. This can help make neural computations more robust, but it can also have an excessive metabolic cost not worth paying. 

    In this paper we have developed a concise yet comprehensive mathematical framework to study the optimality of bilateral versus lateralized solutions. To that end, we have built a series of {\em morphospaces} in which we can look up optimal circuit configurations as a function of (i) costs of running lateralized neural units independently, (ii) costs of coordinating efforts from both brain sides, (iii) how error prone these units are, (iv) the fitness contributed by a successful neural computation, and (v) the complexity of the tasks at hand. 

    A first, very strong result is that only all-or-nothing configurations are optimal within out framework: either it is better to engage both mirror-symmetric systems completely, or just one lateralized half (with the other one permanently shut off), or none at all. It is never exclusively optimal to keep any circuits engaged to an intermediate degree. We further prove that this result is very general (see App.\ \ref{app:5}): it applies to a range of reasonable models that can be conjectured up to weight differently the cost, benefit, fallibility, and complexity dimensions just mentioned. 

    Early findings of localized language function suggested that lateralized brain activity was an exclusive trait of the complex human cognition \cite{Harrington1995, Corballis2009, MacNeilageVallortigara2009}. This was debunked after finding lateralized activity in other animals. However, the hypothesis that cognitive complexity might prompt a break of the brain's mirror symmetry has survived with nuances. A solid mathematical understanding of how such a mechanism might operate has been missing. Our model provides the lacking framework. We show mathematically how different routes to lateralization are fostered by the increasing complexity of computational tasks. We prove how the complexity of strictly emergent phenotypes (which completely displace previously existing simpler ones) can lead to the lateralization of the neural circuitry (Fig.\ \ref{fig:3}{\bf a, e-g}).  We show that this route to lateralization is further favored if both the complex and simple phenotypes are allowed to coexist within the same neural substrate (Fig.\ \ref{fig:4}) and that it is even much more robust if both phenotypes can be allocated each to a different hemisphere (Fig.\ \ref{fig:5}). With these well grounded mathematical results, we conclude that the evolution of more complex cognition can be a paramount driver of brain lateralization. 

    But we also provide strong evidence in the opposite direction: for large combinations of costs, benefits, fallibility, and complexity, there exists a pressure upon formerly lateralized circuits to evolve a duplicate again. The scenarios in which this happens are different from the configurations for which it is optimal to lose ancient bilaterality. Hence, as novel, more complex cognitive phenotypes emerge, they can act as sources of new symmetries or break older, existing ones. Which possibility happens will depend on properties of the neural substrate (e.g.\ its fallibility or metabolic needs) and other conditions (e.g.\ the specific complexity of the emerging phenotype). While we focus our discussion on mirror symmetry, our results are general to any sets of duplicated neural structures. Hence, when a pressure to develop redundancy is present, it might be more parsimonious that it arises within a same hemisphere (by literally duplicating an existing brain area). Evidence of such duplicates of rather large cortical regions in the mammalian brain has recently been described \cite{Seoane2020, ChakrabortyJarvis2015}. The alternative (re-recruiting the actual mirror symmetric units) might be impossible, as they might have been lost or repurposed for other tasks. 

    We visualize optimal configurations as maps (of the model parameters) also called morphospaces. Morphospaces were first introduced to describe the shape of shells as a function of factors affecting their formation \cite{Raup1966, Tyszka2006}. They have been expanded to study other complex systems \cite{Niklas2004, CorominasRodriguez2013, Gonisporns2013, AvenaSporns2015, ArsiwallaVerschure2017, SeoaneSole2018}, including how evolutionary pressures guide the evolution and development of neural or computational substrates \cite{Seoane2019, OlleSole2020, DuongGoni2021}. These maps remind us of phase diagrams that dictate the phase (solid, liquid, etc.)\ of matter samples subjected to different physical conditions. Similarly, the evolutionary pathways that emerge as we move around our morphospaces, or as we superpose the charts of different conditions, remind us of phase transitions. It seems natural to extend these tools and concepts from statistical physics to include computational complexity and reliability (two features closely affected by thermodynamics) -- as we do here. 

    The ultimate goal in the examples of morphospaces just cited is to portray real-world systems and to infer actual phenomenology. Let us try such exercise with our mathematical framework: 
      \begin{itemize}

        \item {\bf Human language} is the most paradigmatic example of lateralization of higher brain function. Language usually involves mostly regions in the left hemisphere \cite{Geschwind1972, CataniFfytche2005, FedorenkoKanwisher2009, FedorenkoKanwisher2012, FedorenkoThompson2014, BlankFedorenko2016, BerwickChomsky2016}, with some symmetric counterparts taking care of prosody or processing non-syntactic patterns \cite{TogaThompson2003}. Recent fMRI evidence shows that response to language starts out as more symmetric in babies, and that it becomes fully lateralized as children grow \cite{OluladeNewport2020}. A similar trajectory can be seen in our model: take neural circuits sitting at the red dots in Figs.\ \ref{fig:4}{\bf e-h} and \ref{fig:5}{\bf e-h}. As language would gradually recruit those neural substrates (i.e.\ as $\tau$ is increased in our model), the initial bilateral configuration becomes suboptimal.

        \item {\bf Hemisphere dominance} is the process through which, while both sides are engaged in some function, one of them takes a much more relevant role -- often acting as a controller of the other or as coordinator of both sides. {\bf Handedness} is a paramount example. There is a strong bias towards mirror symmetry in this case, as the sensory inputs and motor outputs must deal with a bilateral body plan. However, a trend towards laterality is observed in mammals -- with handedness increasing with behavioral complexity \cite{Galaburda1995, RogersVallortigara2004, HalpernRogers2005, Rogers2006}. Patients of unilateral hemiplegia further indicate that the dominant hemisphere is needed (while the dominated one is not) for complex movement of the unaffected hand \cite{Liepmann1905, Harrington1995}. Dominance is observed in many other neural systems such as visual processing \cite{BrownKosslyn1995, Hellige1995} or the {\em theory of mind network} \cite{KliemannTranel2021}. These phenomena might suggest a graded (not all-or-nothing) engagement of the dominated side, which should be a rare configuration according to our mathematical framework. However, our results apply to circuits that are involved in {\em exactly} the same computations. It does not preclude a circuit commanding the other or even delegating specific, unshared tasks on it. Indeed, the different routes to lateralization might promote such controller-controlled configurations. 

        \item {\bf Neural damage} or {\bf pathology} can alter several of the dimensions involved in our model. For example, a damaged circuit will present higher error rates (increased $\varepsilon$) and, potentially, become more costly to engage or coordinate (growing $c$ and $\hat{c}$). Any of these changes would push bilateral circuits outwards from the bilaterality zone (arrows in Fig.\ \ref{fig:2}{\bf c}). {\bf Aging} should also lead to increased fallibility, thus we should expect more asymmetry (as circuits tend to lateralize) with age -- which is the case \cite{ENIGMA2018, ZhouBeaulieu2013, PlessenPeterson2014}. Even if lateralization becomes more optimal due to changes in fallibility or cost, a developed brain might insist on computing with both mirror-symmetric sides -- as if stuck on a frozen accident. If this would happen, it might become helpful to remove one side to achieve the most optimal configuration. A similar conclusion is suggested by a recent model of brain reorganization after hemispherectomy \cite{SeoaneSole2020}. Less invasive treatment (e.g.\ using transcranial magnetic stimulation, TSM, to silence a neural region) might also push suboptimally bilateral circuits in the appropriate direction. 

        \item Larger brains should pay higher coordination costs due to limits on callosal transfer of information. In our model, increased $\hat{c}$ moves mirror-symmetric circuits towards lateralized configurations (vertical arrows in Fig.\ \ref{fig:2}{\bf c}). We hence expect increased asymmetry in larger brains -- as it is the case \cite{ENIGMA2018, KangWoods2015}. This also agrees with evidence that more asymmetric brains present less or thinner fibers across the corpus callosum \cite{Witelson1985} -- thus that they might, perhaps, renounce to some coordination efforts and embrace the lateralized solution. Tasks that demand short reaction times should show similar effects, since they would penalize delays due to inter-hemispheric communication (hence increasing $\hat{c}$). This route to lateralization was already explored in the literature \cite{RingoSimard1994}. Our model subsumes this scenario in a more comprehensive framework. 

        \item Our model can also incorporate parsimoniously other proposed mechanisms for brain lateralization, such as hemisphere specialization \cite{VallortigaraBisazza1999} or optimal packing \cite{Corballis2017} -- both tightly related to our segregated functions from Fig.\ \ref{fig:5}. These proposals were qualitative. We have now built a very general quantitative framework that allows us to understand mathematically how these pathways would work. 

        \item Our morphospaces predict that nearing perfect performance should be yet another pathway to the lateralization of brain function. When $\varepsilon \rightarrow 0$, keeping duplicated circuits is redundant and ineffective (none of our model configurations has bilateral solutions for $\varepsilon = 0$). Musicians with {\bf perfect pitch} provide a case in which this prediction comes true: they have an increased asymmetry in the planum temporale notably owed to the reduction of the non-dominant side for this task \cite{TogaThompson2003, SchlaugSteinmetz1995, Steinmetz1996, KeenanSchlaug2001}. 

      \end{itemize}

    This is a non-exhaustive list of qualitative correspondences between actual neural systems and phenomenology that our morphospaces can explain. Efforts should now be made to bring quantitative empirical results into this theoretical framework. We could try to measure costs, fitness benefits, phenotype complexity, and fallibility in real neural circuits. This seems difficult; but very realistic computer models \cite{Markram2006} might allow us to simulate exact real-world conditions in silico, carefully controlling all dimensions involved. 

    Alternatively, we can try to induce transitions from bilaterality to lateralized solutions in experimental setups, quantify the thresholds at which such changes happen, and, thus, constrain our model parameters. This might be feasible with neural preparations or organoids in vitro, or even in vivo with behaving animals and even humans. For example, we could manipulate task complexity while neural activity is monitored, or we could interfere with the circuitry (e.g.\ using TMS) to raise error rates. We could hope that the brain behaves optimally -- i.e.\ that it will adopt a lateralized mode if it becomes more efficient. But there is no guarantee that this is always the case, so we should also measure performance and energy consumption across individuals to check if those with optimal configurations outperform suboptimal ones. 

    We have tried to make our mathematical framework as general as possible; but, unavoidably, some costs and effects have been left out. Future models should see how our morphospaces change as new aspects come into play. We think that some of the omissions actually make our results more robust. For example, as complex phenotypes emerge, we did not demand that all subtasks are implemented on a same side, and yet we observe clear pathways towards lateralization. Including such penalty to communication across hemispheres should strengthen the trend to asymmetry. We have not discussed structural and lasting costs either: in our model, building a neural circuit is given for granted and we only pay for keeping it running. Such additional costs should exacerbate some of our results -- e.g.\ by making lateralization more definitive, which is relevant for the {\bf L$\boldsymbol{\rightarrow}$B} pathways discussed above. This could introduce path dependency in our evolutionary processes (which reminds us of hysteresis in phase transitions). Other structural constraints might favor bilaterality. We mentioned sensory input and motor output, which are inherently bilateral. It should be feasible to add these conditions to our model and see how the morphospaces are updated. 

    A kind of cost that could alter our results would be non-linearities as a function of the time ($l$ or $r$) that each mirror-symmetric circuit is engaged. We assume that all costs are proportional to $l$ and $r$, mostly because we model an energetic consumption during the time that circuits are engaged. But neurons can be worn out by use, and this process can be highly non-linear -- e.g.\ circuits might break down after a threshold usage, and needed maintenance might saturate with time of use. Including such non-linearities might interfere with some of the mathematical steps in our demonstrations. This does not necessarily change our results, but such models (nonlinear in $l$ and $r$) meed to be studied individually anew. 

    Finally, our model is not only agnostic regarding bilaterality versus other sources of redundancy. They are also independent of the computational substrate. This means that our results should apply when designing efficient computing devices. In such cases (think, a chip that could choose to engage several microprocessors depending on the complexity of the task at hand), it should be helpful to extent our framework to arbitrary numbers of redundant circuits (not just two, as here). A similar modeling might be useful to study gene duplication \cite{HurleyPrince2005, OakleyRivera2008}, especially as more computational and cognitive approaches to the functioning of cells are explored.

\vspace{0.2 cm}

  \section*{Acknowledgments}

    The author wishes to thank Susanna Manrubia for her support in carrying out this research. This work grew through enlightening discussions on brain symmetry and asymmetry with Ricard Sol\'e at the Santa Fe Institute (and elsewhere) and with Susanna Manrubia and her extended group of researchers on Complex Systems at the Spanish National Center for Biotechnology (CNB), the Carlos III University, and other institutions in Madrid, Spain. This work has been funded by the Spanish National Research Council (CSIC) and the Spanish Department for Science and Innovation (MICINN) through a Juan de la Cierva Fellowship (IJC2018-036694-I).

\appendix

  \section{Maxima of the utility function for the simple phenotypes}
    \label{app:1}

    \begin{figure*}[] 
      \begin{center} 
        \includegraphics[width=\textwidth]{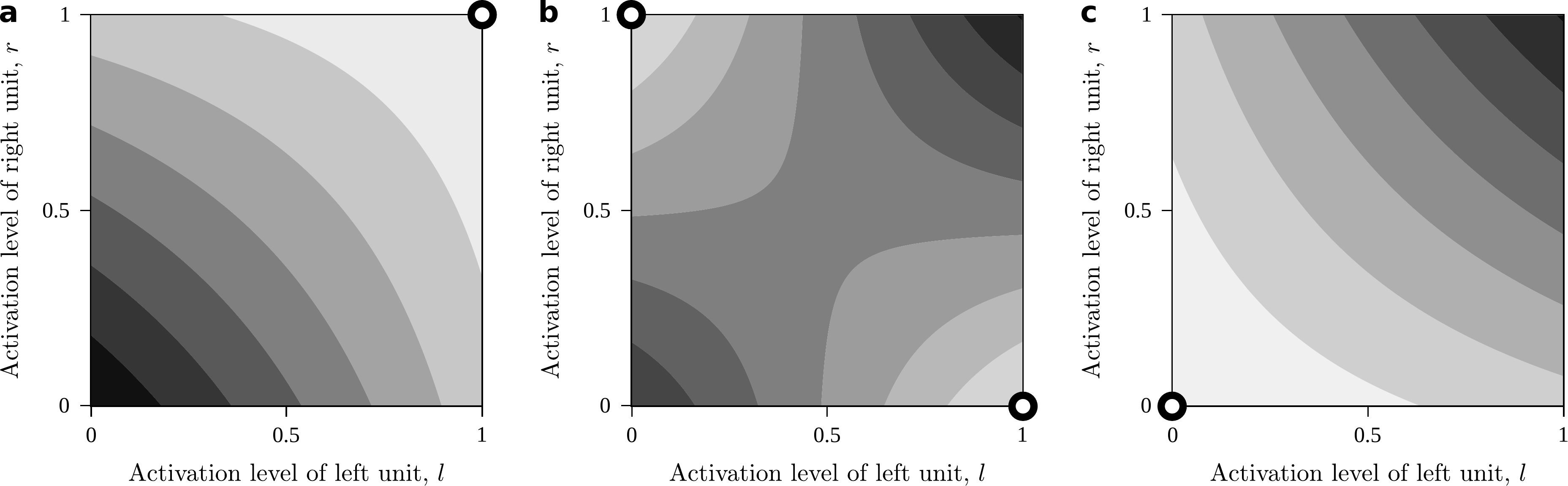}
    
        \caption{{\bf Utility function for the simplest case.} This utility function is a quadratic form with a negative discriminant -- i.e.\ a saddle with no global extrema. Hence, maxima of the utility function within the region of interest are always at the enclosing circuit. With fixed model parameters $b=1$, $c=0.05$, and $\hat{c}=0.1$, varying the error rate we get different optimal solutions (marked by black circles): {\bf a} bilaterality ($\varepsilon = 0.5$), {\bf b} lateralization ($\varepsilon = 0.9$), and {\bf c} no function ($\varepsilon = 0.99$). }
    
        \label{fig:app1.1}
      \end{center}
    \end{figure*}

    In the simplest scenario, the average reward cashed in reads: 
      \begin{eqnarray}
        B = b(1-\varepsilon)\left[ l(1-r) + (1-l)r + (1+\varepsilon)lr \right], 
        \label{eq:app1.1}
      \end{eqnarray}
    and the costs: 
      \begin{eqnarray}
        C &=& c(l+r), \nonumber \\ 
        \hat{C} &=& \hat{c}lr. 
        \label{eq:app1.2}
      \end{eqnarray}
    This yields the utility function: 
      \begin{eqnarray}
        \rho(l,r) &=& b(1-\varepsilon)\left[ l(1-r) + (1-l)r + (1+\varepsilon)lr \right] \nonumber \\ 
        && - c(l+r) - \hat{c}lr. 
        \label{eq:app1.3}
      \end{eqnarray} 
    Given fixed values for our model parameters ($b>0$, $c>0$, $\hat{c}>0$, and $\varepsilon \in [0,1)$), we wish to find maxima of Eq. \ref{eq:app1.3} as a function of our model variables $l \in [0,1]$ and $r \in [0,1]$. Note how we have restricted our parameters: Without loss of generality we could take $b=1$ and normalize all costs accordingly. It makes sense that $c, \hat{c} < b$; otherwise it is never favorable to implement such task (we will see that stronger constraints arise). Finally, the error rate takes values within $\varepsilon \in [0,1]$, but we kept the interval open, $\varepsilon \in [0,1)$, to avoid some algebraic problems. This limit, however, is not so interesting. 

    To start searching for maxima, let us have a look at the derivatives with respect to $l$ and $r$: 
      \begin{eqnarray}
        \rho_l \equiv {\partial \rho \over \partial l} &=& (1-\varepsilon)b-c - r\left[ b(1-\varepsilon)^2 + \hat{c} \right], \nonumber \\
        \rho_r \equiv {\partial \rho \over \partial r} &=& (1-\varepsilon)b-c - l\left[ b(1-\varepsilon)^2 + \hat{c} \right]. 
        \label{eq:app1.4}
      \end{eqnarray}
    These equations are straight lines as a function of $r$ or $l$ respectively. This means that, given fixed model parameters, there is only one point at which both $\rho_l$ and $\rho_r$ can become 0: 
      \begin{eqnarray}
        (l, r) &=& \left( {(1-\varepsilon)b-c \over b(1-\varepsilon)^2 + \hat{c}}, {(1-\varepsilon)b-c \over b(1-\varepsilon)^2 + \hat{c}} \right). 
        \label{eq:app1.5}
      \end{eqnarray}

    Taking second derivatives, 
      \begin{eqnarray}
        \rho_{ll} \equiv {\partial^2 \rho \over \partial l^2} = & 0 & = {\partial^2 \rho \over \partial r^2} \equiv \rho_{rr} 
        \label{eq:app1.6}
      \end{eqnarray}
    and 
      \begin{eqnarray}
        \rho_{lr} \equiv {\partial^2 \rho \over \partial l \partial r} &=& -\left[ b(1-\varepsilon)^2 + \hat{c} \right], 
        \label{eq:app1.7}
      \end{eqnarray}
    yields a discriminant: 
      \begin{eqnarray}
        \Delta &=& \rho_{ll}\rho_{rr} - (\rho_{lr})^2 = -\left[ b(1-\varepsilon)^2 + \hat{c} \right]^2. 
        \label{eq:app1.8}
      \end{eqnarray}
    This number is always negative, meaning that the singular point is a saddle. Since Eq.\ \ref{eq:app1.3} is a quadratic form, the saddle spans the whole $l-r$ plance -- meaning that there are no global or local extrema as a function of our variables. 

    This further implies that, if we look at Eq.\ \ref{eq:app1.3} within a bounded area of the $l-r$ plane, its maxima  will always lie at the boundary (Fig.\ \ref{fig:app1.1}). So we just need to evaluate $\rho(l, r)$ along  the outer circuit of the square $[0,1] \times [0,1]$. Due to the symmetry of the problem, $\rho(l=0,r) = \rho(l,r=0)$ and $\rho(l,r=1) = \rho(l=1,r)$; thus we can just look at $\rho(l, r=0)$ and $\rho(l=1,0)$. 

    \begin{figure*}[] 
      \begin{center} 
        \includegraphics[width=\textwidth]{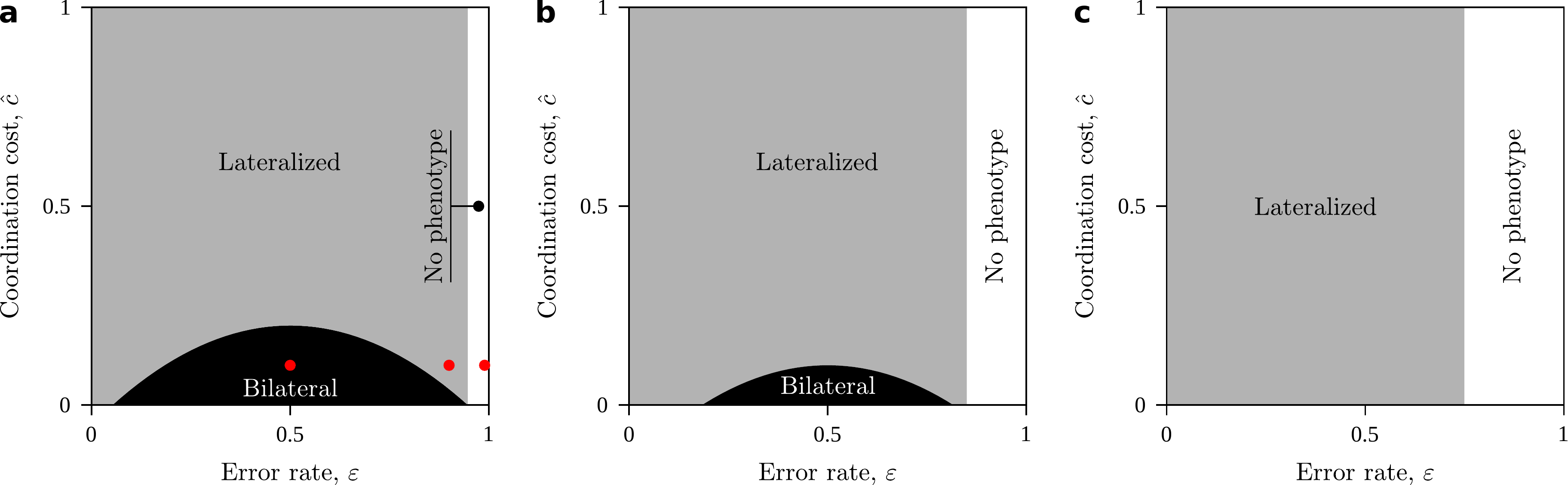}
    
        \caption{{\bf Maps of optimal configurations for simple tasks.} As a function of $\varepsilon$ and $\hat{c}$, a simple phenotype fails to emerge in the white areas because it reports too little fitness given the implementation costs. If the task is worth implementing, it might be so with a lateralized (light gray) or bilateral (black) configuration. {\bf a} Small costs of running independent units, $c = 0.05$. Red dots indicate values of $\varepsilon$ and $\hat{c}$ for which we plot the utility function in Fig.\ \ref{fig:app1.1}. {\bf b} Intermediate values of running independent units, $c = 0.15$. {\bf c} If running independent units is larger than $b/4$ (here $c = 0.25$), the bilateral configuration is not viable. }
    
        \label{fig:app1.2}
      \end{center}
    \end{figure*}

    For the first segment we get: 
      \begin{eqnarray}
        \rho(l, r=0) &=& l\left[ b(1-\varepsilon) - c \right]. 
        \label{eq:app1.9}
      \end{eqnarray}
    As a function of $l$, this is a straight line that passes through the origin. If the slope is positive, the largest value of $\rho(l, r=0)$ will be found at $l=1$. If the slope is negative, the largest value will be found at $l=0$. In such case both neural units are permanently shut off, meaning that it pays off not to implement this phenotype: it is more costly than the reward it produces. We have a negative slope if: 
      \begin{eqnarray}
        c > b(1-\varepsilon) \Rightarrow \varepsilon > 1 - {c \over b}. 
        \label{eq:app1.10}
      \end{eqnarray}
    This condition is marked as a white rectangle in the $\varepsilon-\hat{c}$ maps in Figs.\ \ref{fig:2}{\bf a} and \ref{fig:app1.2}. Precisely at the point when the slope becomes zero, any level of activity $l \in [0, 1]$ is equally optimal. 

    Evaluating the utility function along the second segment of interest we get: 
      \begin{eqnarray}
        \rho(l=1, r) &=& (1-\varepsilon)b - c \nonumber \\ 
        && + r\left[ b(1-\varepsilon)\varepsilon - c - \hat{c} \right], 
        \label{eq:app1.11}
      \end{eqnarray}
    which again is a straight line, now as a function of $r$. Take its slope: 
      \begin{eqnarray}
        m &=& b(1-\varepsilon)\varepsilon - c - \hat{c}. 
        \label{eq:app1.12}
      \end{eqnarray}
    If it is positive, $\rho(l=1, r)$ grows as a function of $r$ and its maxima within $r \in [0,1]$ is found at $r=1$, meaning that it is optimal to keep both neural units active. If the slope is negative, the maxima is found at $r=0$, meaning that it is convenient to keep only one unit (the left one in this case) switched on. Because of the symmetry of the problem, the antisymmetric solution also exists (right unit on and left unit off). If the slope is exactly zero, it is convenient to keep one neural unit always on and it is indistinct whether the other one is on, off, or active at some intermediate level. This happens for: 
      \begin{eqnarray}
        m = 0 &\Leftrightarrow& \hat{c} = -b\varepsilon^2 + b\varepsilon - c. 
        \label{eq:app1.13}
      \end{eqnarray}
    This is a parabola when represented in the $\varepsilon - \hat{c}$ plane. This parabola delimits the regions of parameter space in which it is convenient to switch one or both neural units on (respectively gray and black regions in Figs.\ \ref{fig:2}{\bf a} and \ref{fig:app1.2}). 

    The crossings of the parabola with the horizontal axis are given by imposing $\hat{c} = 0$ on Eq.\ \ref{eq:app1.13}: 
      \begin{eqnarray}
        \varepsilon &=& {1 \pm \sqrt{1 - 4c/b} \over 2}. 
        \label{eq:app1.14}
      \end{eqnarray}
    These crossings are real numbers only if $c < b/4$. Thus, if the individual cost $c$ is larger, a boundary separating bilateral from lateralized optima does not exist -- in that case only the lateralized solution survives (Fig.\ \ref{fig:app1.2}{\bf c}). The maximum of the parabola is $b/4 - c$ and is always located at $\varepsilon=1/2$. This imposes a rather harsh limit on the bilateral configuration: in the best case scenario ($c=0$), coordinating efforts cannot be larger than a fourth of the total fitness contributed by the phenotype -- otherwise, bilateral solutions would not be optimal. Note that the lateralized configuration can incur in much larger costs and still remain viable. 

  \section{Maxima of the utility function for strictly emergent phenotypes}
    \label{app:2} 

    \begin{figure*}[] 
      \begin{center} 
        \includegraphics[width=\textwidth]{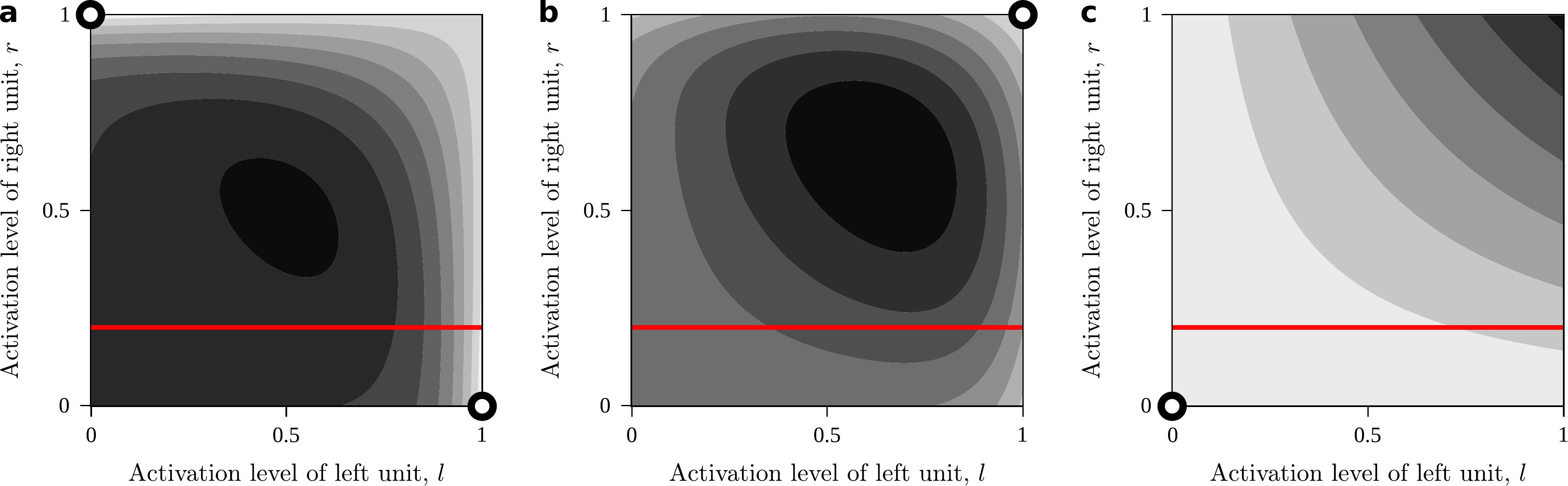}
    
        \caption{{\bf Utility function for the strictly emergent phenotype.} This utility function is a quadratic form with a negative discriminant -- i.e.\ a saddle with no global extrema. Hence, maxima of the utility function within the region of interest are always at the boundary. With fixed model parameters $\tilde{b}=1$, $c=0.05$, and $\hat{c}=0.5$, varying the error rate we get different optimal solutions (marked by black circles): {\bf a} lateralization ($\varepsilon = 0.05$), {\bf b} bilaterality ($\varepsilon = 0.2$), and {\bf c} no function emerges ($\varepsilon = 0.6$). The thick, red line indicates slices of the utility function with constant $r=0.2$ plotted in Fig.\ \ref{fig:app2.2}{\bf b}}
    
        \label{fig:app2.1}
      \end{center}
    \end{figure*}

    Next, we would like to model a complex phenotype that emerges out of the interplay of $K$ individual tasks, each of them solved in an atomic manner by a lateralized or bilateral set of circuits as the ones described in the previous section. Since now we need $K$ such modules to implement the emerging phenotype, assuming that all modules incur in similar costs, we have: 
      \begin{eqnarray}
        C &=& cK(l+r), \nonumber \\ 
        \hat{C} &=& \hat{c}Klr. 
        \label{eq:app2.1}
      \end{eqnarray}
    The fractions $l$ and $r$ can now be interpreted either as the time that modules of a side are active or as the fraction of units of the corresponding side that are always active. 

    Regarding the gained benefit, this is only cashed in if the emergent task is implemented in full, for which we need that all modules work properly. The likelihood that this happens is the product of the likelihood that each of the $K$ units functions correctly: 
      \begin{eqnarray}
        \tilde{B} &=& \tilde{b}K(1-\varepsilon)^K \left[ l(1-r) + (1-l)r + (1+\varepsilon)lr \right]^K. \nonumber \\
        \label{eq:app2.2}
      \end{eqnarray}
    
    Note that here we are modeling a fitness gain brought about strictly by the emergent phenotype -- i.e. we are not considering at this point the benefits from each of the individual tasks (more on that in the next section). We introduce the tilde on $\tilde{B}$ and $\tilde{b}$ to make this explicit. We stipulate that the net fitness gain is $\tilde{b}K$ without loss of generality. We could have absorbed $K$ within $\tilde{b}$ -- but we keep them separated for convenience. It will also be useful to introduce the polynomial: 
      \begin{eqnarray}
        P(l,r) &=& l(1-r) + (1-l)r + (1+\varepsilon)lr \nonumber \\
        &=& r + \left[ 1 - r(1-\varepsilon)\right]l \nonumber \\ 
        &=& l + \left[ 1 - l(1-\varepsilon)\right]r. 
        \label{eq:app2.3}
      \end{eqnarray}
    We have rewritten it three times, the last two just to show that, if $r$ (respectively $l$) are considered constant, the polynomial is a straight line as a function of the other variable. 

    We can now write the following utility function: 
      \begin{eqnarray}
        \rho(l,r) &=& \tilde{B} / K - C / K - \hat{C} / K \nonumber \\ 
        &=& \tilde{b}(1-\varepsilon)^K P(l,r)^K - c(l+r) - \hat{c}lr. 
        \label{eq:app2.4}
      \end{eqnarray}
    Fig.\ \ref{fig:app2.1} shows this function within $(l,r) \in [0,1] \times [0,1]$ in three distinct cases. Again, we are interested in finding maxima of $\rho(l,r)$ within that region of the $l-r$ plane. Therefore let us compute its derivatives with respect to each of the variables: 
      \begin{eqnarray}
        \rho_l \equiv {\partial \rho(l,r) \over \partial l} &=& \tilde{b}(1-\varepsilon)^KKP(l,r)^{K-1}\left[ 1-r(1-\varepsilon) \right] \nonumber \\ 
        && - c - \hat{c}r, \nonumber \\
        \rho_r \equiv {\partial \rho(l,r) \over \partial r} &=& \tilde{b}(1-\varepsilon)^KKP(l,r)^{K-1}\left[ 1-l(1-\varepsilon) \right] \nonumber \\ 
        && - c - \hat{c}r. 
        \label{eq:app2.5}
      \end{eqnarray}

    \begin{figure*}[] 
      \begin{center} 
        \includegraphics[width=\textwidth]{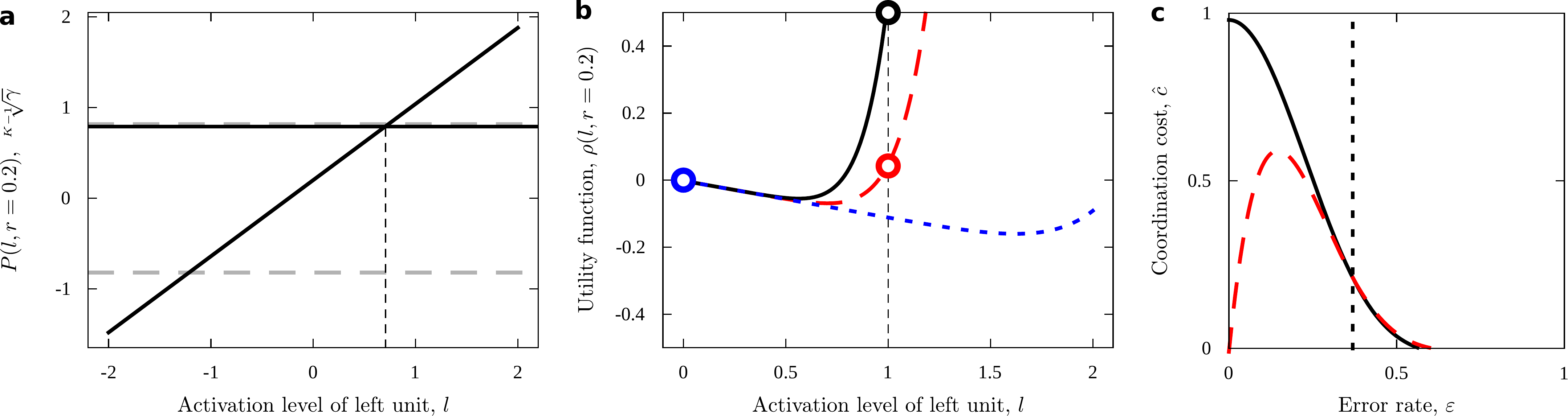}
    
        \caption{{\bf Proving that strictly emergent phenotypes only have all-or-nothing solutions.} {\bf a} The polynomial $P(l,r)$ and the constant $\sqrt[K-1]{\gamma}$ can only cross once, limiting the number of extrema of the utility function. {\bf b} Utility function along lines of constant $r$ from Fig.\ \ref{fig:app2.1}. Their maxima within our range of interest, $l \in [0,1]$, can only be at $l=0$ or $l=1$. {\bf c} Resulting conditions for optimality of either configuration: Dashed vertical line marks when lateralized solutions pay off. The black solid curve indicates when bilateral configurations pay off. The red dashed curve delimits when bilateral configurations are preferred to lateralized ones. }
    
        \label{fig:app2.2}
      \end{center}
    \end{figure*}

    Due to the non-linearity introduced by $K$, this case is more complicated than the one in App.\ \ref{app:1}. To simplify it, let us look at a constant, fixed value of $r$, which defines a straight line parameterized by $l$ in the $l-r$ plane (thick red lines in Fig.\ \ref{fig:app2.1}). Next, let us find singular points along this line: 
      \begin{eqnarray}
        \rho_l = 0 &\Leftrightarrow& P(l,r) = \sqrt[K-1]{c+\hat{c}r \over \tilde{b}(1-\varepsilon)^KK\left[ 1-r(1-\varepsilon) \right]}. \nonumber \\ 
        \label{eq:app2.6}
      \end{eqnarray}
    The right-hand side of this equation is a constant for fixed $r$. Meanwhile, $P(l,r)$ is a straight line as a function of $l$ (as shown by Eq.\ \ref{eq:app2.3}). This straight line can only cross once the constant at the right-hand side (Fig.\ \ref{fig:app2.2}{\bf a}). 

    Let us name the term inside the root as: 
      \begin{eqnarray}
        \gamma &\equiv& {c+\hat{c}r \over \tilde{b}(1-\varepsilon)^KK\left[ 1-r(1-\varepsilon) \right]}. 
        \label{eq:app2.7}
      \end{eqnarray}
    Within the range of parameters and variables that we are interested in, $\gamma$ is always positive, so its $K-1$-th root always exists. If $K$ is even, $K-1$ is odd and there are $K-1$ identical, positive roots at the right-hand side (for $K=10$, black horizontal line in Fig.\ \ref{fig:app2.2}{\bf a}). Thus the equation has $K-1$ identical solutions. If $K$ is odd, $K-1$ is even and there are $(K-1)/2$ identical and negative roots and $(K-1)/2$ identical and positive roots for the right-hand side above (for $K=11$, gray dashed horizontal lines in Fig.\ \ref{fig:app2.2}{\bf a}). Hence, the equation has two sets of $(K-1)/2$ identical solutions -- one set based on the negative root and another one based on the positive root. 

    Introducing the actual dependency of $P(l,r)$ and solving for $l$ we get: 
      \begin{eqnarray}
        l &=& {\sqrt[K-1]{\gamma} - r \over 1-r(1-\varepsilon)}. 
        \label{eq:app2.8}
      \end{eqnarray} 
    In the denominator of this equality we have a positive number (again, for our ranges of parameters and variables), so this does not change the sign of the the solution for $l$. In the numerator, we subtract a positive number ($r>0$ in our range of interest) from $\sqrt[K-1]{\gamma}$. Thus, if we are taking the positive $(K-1)$-th root, we might get a positive or negative solution; and if we are taking the negative root, then we get a negative solution even further away from $0$. 

    Summarizing,  Eq.\ \ref{eq:app2.6} might have either one positive and one negative, or two negative solutions for our range of parameters. This means that the derivative of the utility function changes sign at most once for $l \ge 0$, thus it has at most one extremum for positive $l$. 

    For $l\rightarrow \infty$, the utility function is dominated by: 
      \begin{eqnarray}
         P(l,r)^K &=& \left[ 1 - r(1-\varepsilon) \right]^K l^K, 
        \label{eq:app2.8}
      \end{eqnarray} 
    which is positive and growing as a function of $l$. Thus, if there is one extremum for $l>0$, it must be a minimum. This means that the maximum of the utility function along a line with fixed $r$ and $l\in[0,1]$ must be found either at $l=0$ or $l=1$ (Fig.\ \ref{fig:app2.2}{\bf b}). 
      
    But this reasoning is valid for any constant $r$ within our range of interest ($r\in[0,1]$) and also, symmetrically, for any constant $l\in[0,1]$ if we take the utility function as depending on $r$ alone. This means that the maxima of our utility function for $(l,r) \in [0,1]\times [0,1]$ must be found around the contour and, specifically, either at $(l,r)=(0,0)$, $(l,r)=(1,0)$, $(l,r)=(0,1)$, or $(l,r)=(1,1)$. Due to symmetry, if there is a maximum at $(l,r)=(1,0)$, there is another one at $(l,r)=(0,1)$. Thus we only need to compare three points to solve our problem: 
      \begin{eqnarray}
        \rho(0,0) &=& 0. \nonumber \\ 
        \rho(1,0) &=& \tilde{b}(1-\varepsilon)^K - c. \nonumber \\
        \rho(1,1) &=& \tilde{b}(1-\varepsilon)^K(1+\varepsilon)^K - 2c - \hat{c} \nonumber \\
        &=& \tilde{b}(1-\varepsilon^2)^K - 2c - \hat{c}. 
        \label{eq:app2.9}
      \end{eqnarray}

    Imposing that the fully lateralized solution ever pays off, $\rho(1,0)>0$, yields: 
      \begin{eqnarray}
        c &<& \tilde{b}(1-\varepsilon)^K \nonumber \\ 
        &\Rightarrow& \varepsilon < 1 - \sqrt[K]{c \over \tilde{b}}, 
        \label{eq:app2.10}
      \end{eqnarray} 
    which traces a straight vertical line in the $\varepsilon - \hat{c}$ map (dotted black line in Fig.\ \ref{fig:app2.2}{\bf c}). 

    Imposing that the bilateral solution ever pays off ($\rho(1,1)>0$) yields: 
      \begin{eqnarray}
        \hat{c} &<& \tilde{b}(1-\varepsilon^2)^K - 2c. 
        \label{eq:app2.11}
      \end{eqnarray}
    This is a non-linear, monotonically decreasing function of $\varepsilon$ (solid black curve in Fig.\ \ref{fig:app2.2}{\bf c}). 

    Finally, we check when the bilateral solution is preferred to the lateralized one ($\rho(1,1) > \rho(1,0)$): 
      \begin{eqnarray}
        \hat{c} &<& \tilde{b}(1-\varepsilon)^K\left[ (1+\varepsilon)^K - 1 \right] - c. 
        \label{eq:app2.12}
      \end{eqnarray} 
    This is another non-linear curve on the $\varepsilon - \hat{c}$ plane, with a unique maximum (dashed red line in Fig.\ \ref{fig:app2.2}{\bf c}). 

    The interplay between these conditions yields a map that tells us whether the phenotype is too costly to emerge (large white area in Fig.\ \ref{fig:2}{\bf b}), or whether fully lateralized (gray area in Fig.\ \ref{fig:2}{\bf b}) or bilateral (black in Fig.\ \ref{fig:2}{\bf b}) neural structures are preferred. We can compute that the conditions given by Eqs.\ \ref{eq:app2.11} and \ref{eq:app2.12} cross precisely at $\varepsilon = 1 - \sqrt[K]{c / \tilde{b}}$. This means that, while Eq.\ \ref{eq:app2.12} describes the bilateral region for $\varepsilon < 1 - \sqrt[K]{c / \tilde{b}}$, the rightmost part of the bilateral region is given by Eq.\ \ref{eq:app2.11}.

  \section{Maxima for emerging phenotype along previously existing, simpler ones}
    \label{app:3} 
      
    \begin{figure}[] 
      \begin{center} 
        \includegraphics[width=0.9\columnwidth]{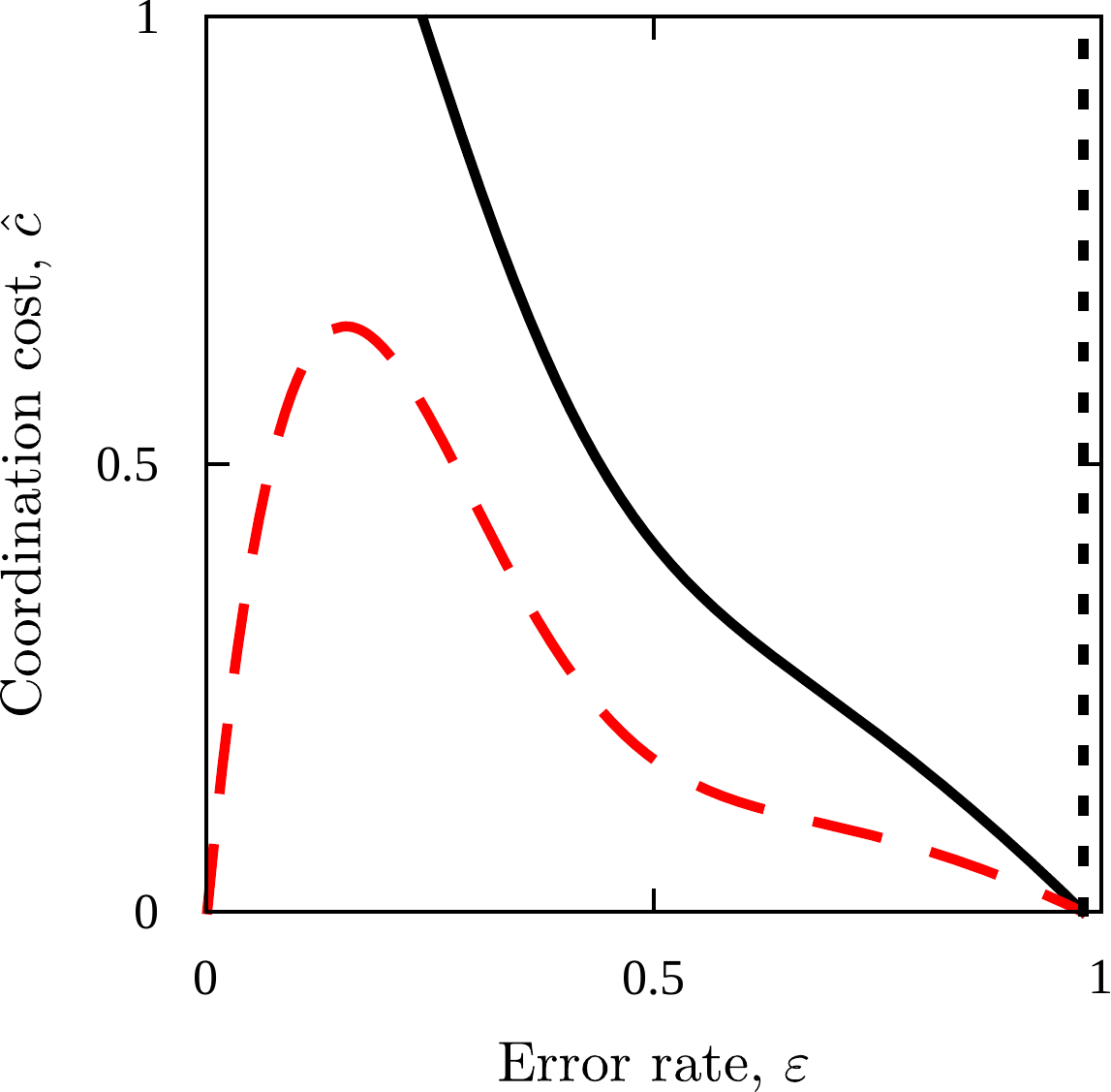}
    
        \caption{{\bf Resulting conditions for optimal configurations in phenotypes that emerge along earlier, simpler tasks.} The vertical dashed line indicates a lower bound for phenotype viability. For $\varepsilon$ at the left of this line, the emergent phenotype (along with implementation of the earlier, simpler tasks) is viable. At the right of this line, the phenotype might not be viable (but we cannot rule it out since Eq.\ \ref{eq:app3.7} is an approximation). The black solid curve indicates when bilateral configurations pay off. The red dashed curve delimits when bilateral configurations are preferred to lateralized ones. For large $K$, these curves do not cross, meaning that the red dashed curve is a more stringent condition for bilaterality. For smaller $K$, they might cross (and, if so, they allow the evolutionary pathway {\bf \O$\boldsymbol{\rightarrow}$B} from Fig.\ \ref{fig:3}{\bf e}). }
    
        \label{fig:app3.1}
      \end{center}
    \end{figure}

    In the previous section we studied an emergent phenotype that contributed some fitness only if all building blocks (i.e.\ all individual tasks) were successfully implemented. However, since such emergent tasks evolve upon a previously existing substrate, it might be the case that each individual neural module contributes some fitness of its own irrespective of the emergent task -- e.g.\ because they still carry out their ancient task during a fraction of the time. To model this, we add up the benefits reported both by the acient and the emergent phenotypes: 
      \begin{eqnarray}
        \tilde{B} + B &=& \tilde{b}K(1-\varepsilon)^KP(l,r)^K + bK(1-\varepsilon)P(l,r), \nonumber \\ 
        \label{eq:app3.1}
      \end{eqnarray}
    where $P(l,r)$ is as before. We assume that the costs still depend on the engagement of each neural unit, so they are as in Eq.\ \ref{eq:app2.1}. We can write the following utility function: 
      \begin{eqnarray}
        \rho(l,r) &=& \tilde{B}/K + KB/K + C/K + \hat{C}/K \nonumber \\
        &=& \tilde{b}(1-\varepsilon)^KP(l,r)^K + b(1-\varepsilon)P(l,r) \nonumber \\ 
        && -c(l+r) - \hat{c}lr. 
        \label{eq:app3.2}
      \end{eqnarray}

    As usual, we need to find maxima of $\rho(l,r)$ within $(l,r) \in [0,1]\times[0,1]$. Let us look again at derivatives and singular points along a straight line of constant, fixed $r$ and as a function of $l$: 
      \begin{eqnarray}
        \rho_l \equiv {\partial \rho(l,r) \over \partial l} &=& \tilde{b}(1-\varepsilon)^KKP(l,r)^{K-1}\left[ 1 - r(1-\varepsilon) \right] \nonumber \\ 
        && + b(1-\varepsilon)\left[ 1 - r(1-\varepsilon) \right] - c - \hat{c}r. \nonumber \\
        \label{eq:app3.3}
      \end{eqnarray}
    This becomes zero if: 
      \begin{eqnarray}
        \rho_l=0 &\Leftrightarrow& P(l,r) = \sqrt[K-1]{ {c+\hat{c} \over \tilde{b}(1-\varepsilon)K \left[ 1 - r(1-\varepsilon) \right]} - {b\over \tilde{b}K} } \nonumber \\ 
        && = \sqrt[K-1]{\gamma - {b\over \tilde{b}K}}. 
        \label{eq:app3.4}
      \end{eqnarray}
    
    We can apply here the same argument as before (illustrated in Fig.\ \ref{fig:app2.2}{\bf a-b}). At the left-hand side of this equation we have a straight line as a function of $l$. At the right-hand side, we have, depending on whether $K-1$ is odd or even, either one or two constant values as a function of $l$. The negative root for even $K-1$ is not interesting for us. The positive one for even $K-1$ or for odd $K-1$ might result in a solution of Eq.\ \ref{eq:app3.4} for positive $l$ within $l\in[0,1]$. If this singular point is an extremum of the utility function, it must be a minimum. The right-hand side of Eq.\ \ref{eq:app3.4} is the same as the right-hand side of Eq. \ref{eq:app2.6} with a small correction, $\tilde{b}/bK$. This correction might turn the argument of the root negative, thus there would not be any root in the real numbers for even $K-1$. This does not affect our argument, as it just means that extrema do not exist for $l \in [0,1]$ in such a case. 

    All this reasoning implies that the utility function with fixed $r$ and taking values over $l\in[0,1]$ has its maximum either at $l=0$ or $l=1$. Thus, again, within out region of interest, Eq.\ \ref{eq:app3.2} can only have its maxima at $(l,r)=(0,0)$, $(l,r)=(1,0)$, $(l,r)=(0,1)$, or $(l,r)=(1,1)$. And, again, $(l,r)=(1,0)$ and $(l,r)= (0,1)$ will be maxima simultaneously. We just need to evaluate three points: 
      \begin{eqnarray}
        \rho(0,0) &=& 0. \nonumber \\ 
        \rho(1,0) &=& \tilde{b}(1-\varepsilon)^K + b(1-\varepsilon) - c. \nonumber \\ 
        \rho(1,1) &=& \tilde{b}(1-\varepsilon^2)^K + b(1-\varepsilon^2) - 2c - \hat{c}. 
        \label{eq:app3.5}
      \end{eqnarray}

    Imposing that the fully lateralized solution is larger than $0$ we get: 
      \begin{eqnarray}
        \rho(1,0)>0 &\Leftrightarrow& c < \tilde{b}(1-\varepsilon)^K + b(1-\varepsilon). 
        \label{eq:app3.6}
      \end{eqnarray}
    The right-hand side is dominated by the second term when $K$ is large. In any case, we can be sure that this condition is met if we dismiss the first term and attend to the constraint: 
      \begin{eqnarray}
        \varepsilon &<& 1-c/b. 
        \label{eq:app3.7}
      \end{eqnarray}
    
    Imposing that the bilateral solution has positive utility we get: 
      \begin{eqnarray}
        \rho(1,1) > 0 &\Leftrightarrow& \hat{c} < \tilde{b}(1-\varepsilon^2)^K + b(1-\varepsilon^2) - 2c. \nonumber \\
        \label{eq:app3.8}
      \end{eqnarray} 
    This is depicted as a solid black curve in Fig.\ \ref{fig:app3.1}. Finally, we check when the bilateral solution is preferred to the fully lateralized one, $\rho(1,1)>\rho(1,0)$: 
      \begin{eqnarray}
        \hat{c} &<& \tilde{b}(1-\varepsilon)^K\left[ (1+\varepsilon)^K - 1 \right] + b\varepsilon(1-\varepsilon) - c. \nonumber \\
        \label{eq:app3.9}
      \end{eqnarray}
    This is the dashed red curve in Fig.\ \ref{fig:app3.1}. As before, we can study when these two conditions cross. Imposing that Eq.\ \ref{eq:app3.8} equals Eq. \ref{eq:app3.9} we get: 
      \begin{eqnarray}
        \tilde{b}x^{K-1} + b &=& {c\over x}, 
        \label{eq:app3.10}
      \end{eqnarray}
    with the change of variable $\varepsilon \rightarrow 1-x \Rightarrow x = 1 - \varepsilon$. Eq.\ \ref{eq:app3.10} is a monotonously increasing polynomial that crosses a hyperbola only once, and this crossing happens for smaller $x$ the larger the $K$. This means that, as more neural units are involved, both conditions cross for $\varepsilon \rightarrow 1$, thus Eq.\ \ref{eq:app3.8} is the most relevant for us for large $K$. 

    Again, the interplay between these conditions renders a phase space where we can read what configuration of neural circuits (lateralized, bilateral, or none) is preferred depending on the model parameters (Fig.\ \ref{fig:2}{\bf c}). All the conditions just derived are basically the same as in the previous section plus some correction contributed by the non-emergent phenotypes. This contribution enables both neural configurations in a wider range of model parameters. Notably, the region in which the phenotypes become unfeasible is drastically reduced. 

  \section{Segregating functions}
    \label{app:4}

    The emergence of complex phenotypes upon a neural circuitry that keeps implementing more ancient tasks opens up the possibility of segregating both computations -- thus lateralizing and confining each phenotype to a hemisphere. We check when such a solution is favorable as compared to the fully lateralized and mirror-symmetric configurations described in the previous appendix. 

    Say that the emergent phenotype contributes some fitness $\tilde{b}'$ when it is implemented, which happens a fraction $\tau$ of the time -- resulting in $\tilde{b} = \tau\tilde{b}'$. Similarly, each ancient, simple task contributes a fitness $b'$ during the remaining fraction $1-\tau$ of the time -- resulting in $b=(1-\tau)b'$. From Eq.\ \ref{eq:app3.5}, the resulting utility functions for the fully lateralized and bilateral solutions read: 
      \begin{eqnarray}
        \rho(1,0) &=& \tau\tilde{b}'(1-\varepsilon)^K + (1-\tau)b'(1-\varepsilon) - c, \nonumber \\ 
        \rho(1,1) &=& \tau\tilde{b}'(1-\varepsilon^2)^K + (1-\tau)b'(1-\varepsilon^2) - 2c - \hat{c}. \nonumber \\
        \label{eq:app4.1}
      \end{eqnarray}

    In these configurations, including the fully lateralized one, all functions are performed in a same circuit. Instead, segregating functionality allows two sets of circuits: one for the emergent phenotype and another one for the simpler tasks. We assume, incidentally, that the complex phenotype displaces simpler tasks in the unsegregated solution a fraction $\tau$ of the time -- hence their contribution $ (1-\tau)$. Now that phenotypes are segregated, simpler tasks can make their full contribution again. With this in mind, we compute the costs and benefits of function segregation. Say we lateralize the complex phenotype to the left and the simpler tasks to the right, and we note this $\left<L\right|$ and $\left|R\right>$ respectively. Then: 
      \begin{eqnarray}
        \rho(\left<L|R\right>) &=& \tau\tilde{b}'(1-\varepsilon)^K + b'(1-\varepsilon) - (1+\tau)c. \nonumber \\
        \label{eq:app4.2}
      \end{eqnarray}
    Here we have used that $P(1,0) = 0 = P(0,1)$, and that there are no coordination costs across hemispheres if function is segregated. 

    First we find that $\rho(\left<L|R\right>) > \rho(1,0)$ for any $\varepsilon < 1 - c/b'$. This includes any configuration in which lateralization is preferred, meaning that whenever lateralization is optimal, it is even more optimal to segregate function and engage each hemisphere with a different task. This also includes configurations in which $\rho(1,0) < 0$ (i.e.\ in which a lateralized circuit fails to be cost-efficient enough as to be implemented). In that region, is the segregated circuit worth it? We find that $\rho(\left<L|R\right>) > 0$ for: 
      \begin{eqnarray}
        (1+\tau) c &<& \tau\tilde{b}'(1-\varepsilon)^K + b'(1-\varepsilon). 
        \label{eq:app4.3}
      \end{eqnarray}
    To approximate this inequality roughly, as above, we ignore the first term at the right hand side (which becomes small for large $K$), finding: $\varepsilon < 1 - (1+\tau)c/b'$. 

    Finally, we compare $\rho(\left<L|R\right>)$ and $\rho(1,1)$ to find when does the bilateral configuration remain optimal: 
      \begin{eqnarray}
        \hat{c} &<& \tau\tilde{b}'(1-\varepsilon)^K\left[ (1+\varepsilon)^K - 1 \right] \nonumber \\ 
        && + b'(1-\varepsilon)\left[ \varepsilon - \tau - \tau\varepsilon \right] - c(1-\tau). 
        \label{eq:app4.4}
      \end{eqnarray}

    The interplay between these curves renders maps of optimality of bilateral versus lateralized and segregated configurations. It is noteworthy that a region of parameters exist (as given by Eq.\ \ref{eq:app4.4}) in which the bilateral configuration pays off. This means that the advantage of redundant computations can overcome the added fitness gain from both, segregated phenotypes.

  \section{Generality of results} 
    \label{app:5} 

    One strong assumption above is that the fitness gained is an all or nothing function of the implemented task. Either
    a computation is implemented perfectly (which happens with probability $1-\varepsilon$ in each individual circuit)
    or it is not implemented at all. But we could reinterpret $\varepsilon$ as a measure of a distance between the
    target computation and the output generated by the circuit, and assume that some fitness is still obtained if this
    distance is acceptable. Might some of our results be more general if we lift that constraint? 

    Let us assume that each individual unit contributes a fitness that is a generic, non-linear function $f
    (\varepsilon)$. Similarly, if both units are engaged, they contribute with another function $g(\varepsilon)$ that
    should capture that, working together, a higher accuracy might be reached. It makes sense that $f(\varepsilon)$ and
    $g(\varepsilon)$ are monotonously decreasing with $\varepsilon$. 

    On the other hand, we can impose generic costs $h(\varepsilon)$ and $\hat{h}(\varepsilon)$. We can demand that these
    costs increase if each unit independently (or both together) is required to function with greater accuracy. This
    would entail large costs for low $\varepsilon$ and vice-versa -- thus $h(\varepsilon)$ and $\hat{h}
    (\varepsilon)$ should also be positive, monotonously decreasing with $\varepsilon$, and non-linear in general. 

    For the simplest case of a single, non-emergent phenotype, we get: 
      \begin{eqnarray}
        \rho(l,r) &=& b\left[ f(\varepsilon)l(1-r) + f(\varepsilon)(1-l)r + g(\varepsilon)lr \right] \nonumber \\
        &=& -ch(\varepsilon)(l+r) - \hat{c}\hat{h}(\varepsilon)lr, 
        \label{eq:app5.1}
      \end{eqnarray}
    with derivatives: 
      \begin{eqnarray}
        \rho_l \equiv {\partial \rho(l,r) \over \partial l} &=& b\left[ f(\varepsilon)(1-2r)+g(\varepsilon)r \right] \nonumber \\
        && - ch(\varepsilon) - \hat{c}\hat{h}(\varepsilon)r, \nonumber \\
        \rho_r \equiv {\partial \rho(l,r) \over \partial r} &=& b\left[ f(\varepsilon)(1-2l)+g(\varepsilon)l \right] \nonumber \\
        && - ch(\varepsilon) - \hat{c}\hat{h}(\varepsilon)l; 
        \label{eq:app5.2}
      \end{eqnarray}
    and second derivatives: 
      \begin{eqnarray}
        \rho_{ll} \equiv {\partial^2 \rho(l,r) \over \partial l^2} = & 0 & = {\partial^2 \rho(l,r) \over \partial r^2} = \rho_{rr}, \nonumber \\
        \rho_{lr} \equiv {\partial^2 \rho(l,r) \over \partial l \partial r} &=& b\left[ g(\varepsilon)-2f(\varepsilon) \right] - \hat{c}\hat{h}(\varepsilon). 
        \label{eq:app5.3}
      \end{eqnarray}
    However complicated $f(\varepsilon)$, $g(\varepsilon)$, or $\hat{h}(\varepsilon)$ might be, for constant values of
    $\varepsilon$ the utility function is always a quadratic form and its discriminant is always negative: 
      \begin{eqnarray}
        \Delta &\equiv& \rho_{ll}\rho_{rr} - \left(\rho_{lr}\right)^2 < 0. 
        \label{eq:app5.4}
      \end{eqnarray}
    As we saw above, this implies that $\rho(l,r)$ is always a saddle. As such, there cannot be any extrema within
    enclosed areas, and maxima of the utility function will lie along the enclosing circuit. 

    As before, we note that $\rho(l, r=0)$ is equivalent to $\rho(l=0, r)$ and that $\rho(l=1, r)$ is equivalent to
    $\rho(l, r=1)$; thus we only need to evaluate two segments. Along the first one we get: 
      \begin{eqnarray}
        \rho(l, r=0) &=& \left[ bf(\varepsilon) - ch(\varepsilon) \right]l, 
        \label{eq:app5.5}
      \end{eqnarray}
    and along the second one we get: 
      \begin{eqnarray}
        \rho(l=1, r) &=& bf(\varepsilon) - ch(\varepsilon) \nonumber \\
        && + \left[ b(g(\varepsilon) - f(\varepsilon)) - ch(\varepsilon) -\hat{c}\hat{h}(\varepsilon) \right]r. \nonumber \\
        \label{eq:app5.6}
      \end{eqnarray}

    Both these expressions are straight lines as a function of the independent variable ($l$ and $r$ respectively). This
    means that, unless when the slope becomes $0$, the maxima are found either at $(l,r)=(0,0)$, at $(l,r)=(1,0)$, or
    at $(l,r)=(1,1)$. Thus again we only need to check out the corners of the enclosed area. When the slope along
    either segment becomes $0$, the whole segment has the same utility function and the maximum along the circuit might
    become degenerate. In this case we would have as solution a range of activity for either neural unit -- while the
    other would still be either completely engaged or shut off. Depending on the functional form of $f
    (\varepsilon)$, $g(\varepsilon)$, $g(\varepsilon)$, and $\hat{h}(\varepsilon)$, this might happen several countable
    times as a function of $\varepsilon$. But, still, an uncountable number of times the function will have its maxima
    in the corners of $(l, r) \in [0,1] \times [0,1]$. 

    This analysis means that, in general, fully lateralized and completely bilateral solutions are the only
    evolutionarily stable configurations for non-emergent tasks. This result does not depend on the specific forms of
    $f(\varepsilon)$, $g(\varepsilon)$, $h(\varepsilon)$, and $\hat{h}(\varepsilon)$. Note that, so far, we have not
    even used the fact that these functions are monotonously decreasing. \\

    Can we generalize this result to emergent phenotypes? Let us move straightaway to the most complete case, in which
    we have emergent phenotypes together with the individual ones for each unit. The utility function now reads: 
      \begin{eqnarray}
        \rho(l,r) &=& \tilde{b}\tilde{P}(l,r)^K + bP(l,r) - ch(\varepsilon) - \hat{c}\hat{h}(\varepsilon)lr, \nonumber \\
        \label{eq:app5.7}
      \end{eqnarray}
    where: 
      \begin{eqnarray}
        \tilde{P}(l,r) &=& \tilde{f}(\varepsilon)l(1-r) + \tilde{f}(\varepsilon)(1-l)r + \tilde{g}(\varepsilon)lr = \nonumber \\ 
        &=& l\left[ (1-2r)\tilde{f}(\varepsilon) + r\tilde{g}(\varepsilon) \right] + r\tilde{f}(\varepsilon) = \nonumber \\
        &=& r\left[ (1-2r)\tilde{f}(\varepsilon) + l\tilde{g}(\varepsilon) \right] + l\tilde{f}(\varepsilon). 
        \label{eq:app5.8}
      \end{eqnarray}
    As before, we have written this polynomial in three different ways to show that the polynomial is a straight line of
    $l$ (respectively of $r$) if we take a fixed, constant $r$ (respectively $l$). This will be useful. A similar
    polynomial $P(l,r)$ is introduced in the utility function for the non-emergent phenotype (with its corresponding $f
    (\varepsilon)$ and $g(\varepsilon)$, not necessarily similar to $\tilde{f}(\varepsilon)$ and $\tilde{g}
    (\varepsilon)$). 

    Also as before, let us look at the utility function along a line of fixed, constant $r$ and as a function of $l$. To
    search for extrema along such a line we look at the derivative: 
      \begin{eqnarray}
        \rho_l \equiv {\partial \rho(l, r)\over \partial l} &=& \tilde{b}K\tilde{P}^{K-1}(l,r)\left[ (1-2r)\tilde{f}(\varepsilon) + r\tilde{g}(\varepsilon) \right] \nonumber \\ 
        && + b\left[ (1-2r)f(\varepsilon) + rg(\varepsilon) \right] \nonumber \\
        && - ch(\varepsilon) - \hat{c}\hat{h}(\varepsilon)r. 
        \label{eq:app5.9}
      \end{eqnarray}
    So we get: 
      \begin{eqnarray}
        \rho_l=0 &\Leftrightarrow& \tilde{P}(l,r) = \sqrt[K-1]{\gamma}, \nonumber \\ 
        \gamma &\equiv& ch(\varepsilon) + \hat{c}\hat{h}(\varepsilon)r - b\left[ (1-2r)f(\varepsilon) + rg(\varepsilon) \right] \over \tilde{b}K\left[ (1-2r)\tilde{f}(\varepsilon) + r\tilde{g}(\varepsilon) \right]. \nonumber \\ 
        \label{eq:app5.10}
      \end{eqnarray}

    Again we have solutions to this equation when a straight line as a function of $l$ crosses either one or two
    constant values (as illustrated in Fig.\ \ref{fig:app2.2}{\bf a}). Introducing the actual form of $\tilde{P}(l,r)$ we get: 
      \begin{eqnarray}
        l &=& { \sqrt[K-1]{\gamma} - r\tilde{f}(\varepsilon) \over  (1 - 2r)\tilde{f}(\varepsilon) + r\tilde{g}(\varepsilon)}. 
        \label{eq:app5.11}
      \end{eqnarray}
    In the numerator we have either the positive or negative $(K-1)$-th root of $\gamma$ minus $r\tilde{f}
    (\varepsilon)$, a positive term. If $K-1$ is odd we get either one negative or one positive value in the numerator.
    If $K-1$ is even and $r\tilde{f}(\varepsilon) < ||\sqrt[K-1]{\gamma}||$ we get one positive and one negative
    values. If $K-1$ is even and $r\tilde{f}(\varepsilon) > ||\sqrt[K-1]{\gamma}||$ we get two negative values in the
    numerator. If this is ever the case, and the denominator is negative, the two negative values in the numerator
    could turn into two singular points within $l \in[0,1]$, and one of them might be a maximum. In any other
    situation, again, maxima of $\rho(l,r)$ will lie either at $(l,r)=(0,0)$, $(l,r)=(1,0)$, $(l,r)=(0,1)$, $(l,r)=
    (1,1)$. 

    When do we get a negative number in the denominator? 
      \begin{eqnarray}
        (1 - 2r)\tilde{f}(\varepsilon) + r\tilde{g}(\varepsilon) < 0 &\Leftrightarrow& {\tilde{g}(\varepsilon) \over \tilde{f}(\varepsilon)} > 2 - {1 \over r}. \nonumber \\
        \label{eq:app5.12}
      \end{eqnarray}
    In the simple model studied above we have chosen: $\tilde{f}(\varepsilon) = 1 - \varepsilon$ and $\tilde{g}
    (\varepsilon) = 1-\varepsilon^2$. We get: 
      \begin{eqnarray}
        {\tilde{g}(\varepsilon) \over \tilde{f}(\varepsilon)} &=& {(1-\varepsilon)(1+\varepsilon) \over 1-\varepsilon} = 1+\varepsilon. 
        \label{eq: app4.13}
      \end{eqnarray}
    This number is always larger than $1$ and, within our range of interes ($r \in [0,1]$), $2-1/r$ is always greater
    than $1$. This is, our chosen functions fullfil the condition and never have graded activities of the circuits, as
    shown above by other means. 

    In general, we can request $\tilde{f}(\varepsilon)$ and $\tilde{g}(\varepsilon)$ to behave as $\tilde{f}
    (\varepsilon) = (1-\varepsilon)^\alpha$ and $\tilde{g}(\varepsilon) = (1-\varepsilon)^\beta$. Note that $2 - 1/r$
    is always negative for $r<1/2$, and that $\tilde{g}/\tilde{f}$ is always positive because it is the quotient of two
    positive numbers. Thus functions of this form are not problematic for $r<1/2$. In the wors-case scenario
    ($r\rightarrow1$, thus $2-1/r \rightarrow 1$), the quotient $\tilde{g}(\varepsilon)/\tilde{f}(\varepsilon)$ will
    remain larger than $1$ if $\tilde{g}(\varepsilon) > \tilde{f}(\varepsilon)$, which is always the case if $\beta
    < \alpha$. An exponent $\beta < \alpha$ implies that the function $\tilde{g}(\varepsilon)$ decreases with
    $\varepsilon \in [0,1)$ more slowly than $\tilde{f}(\varepsilon)$. I.e., this condition is requesting that the
    payoff attained by two circuits working together is higher than that of one circuit working on its own -- which is
    a reasonable request to make. 

    All in all we can confirm that, for ample, reasonable assumptions, configurations of neural circuits of the kind
    studied here only have all-or-nothing solutions. This means, we either observe fully lateralized or completely
    bilateral configurations, and we should not observe a unit engaged only during a fraction of the time if both are
    implementing precisely the same computations.

\end{document}